\documentclass[nofootinbib,twocolumn,unsortedaddress,superscriptaddress]{revtex4}
\usepackage{amsmath}
\usepackage{amssymb}
\usepackage{graphicx}
\usepackage{color}
\usepackage{subfigure}
\usepackage{multirow}
\newcommand{\vect}[1]{\boldsymbol #1}
\newcommand{\mat}[1]{\boldsymbol #1}
\makeatother
\date{\today}
\newcommand{\nn}{\nonumber\\}

\begin{document}
\title{Model Cortical Association Fields Account for the Time Course and Dependence on Target Complexity of Human Contour Perception}

\author{Vadas Gintautas}
\email{vgintautas@chatham.edu}
\affiliation{Center for Nonlinear Studies and T-5, Theoretical Division, Los Alamos National Laboratory, Los Alamos, New Mexico, USA}
\affiliation{Physics Department, Chatham University, Pittsburgh, Pennsylvania, USA}
\author{Michael I. Ham}
\affiliation{P-21, Physics Division, Los Alamos National Laboratory, Los Alamos, New Mexico, USA}
\author{Benjamin Kunsberg}
\author{Shawn Barr}
\affiliation{New Mexico Consortium, Los Alamos, New Mexico, USA}
\author{Steven P. Brumby}
\affiliation{Space and Remote Sensing Sciences, Los Alamos National Laboratory, Los Alamos, New Mexico, USA}
\author{Craig Rasmussen}
\affiliation{New Mexico Consortium, Los Alamos, New Mexico, USA}
\author{John S. George}
\affiliation{P-21 Applied Modern Physics (Biological and Quantum Physics), Los Alamos National Laboratory, Los Alamos, New Mexico, USA}
\author{Ilya Nemenman}
\affiliation{Departments of Physics and Biology and Computational and Life Sciences Initiative, Emory University, Atlanta, Georgia, USA}
\author{Lu{\'i}s M. A. Bettencourt}
\affiliation{Center for Nonlinear Studies and T-5, Theoretical Division, Los Alamos National Laboratory, Los Alamos, New Mexico, USA}
\author{Garrett T. Kenyon}
\email{garkenyon@gmail.com}
\affiliation{New Mexico Consortium, Los Alamos, New Mexico, USA}

\maketitle
\section*{Abstract} 
Can lateral connectivity in the primary visual cortex account for the time dependence and intrinsic task difficulty of human contour detection? To answer this question, we created a synthetic image set that prevents sole reliance on either low-level visual features or high-level context for the detection of target objects. Rendered images consist of smoothly varying, globally aligned contour fragments (amoebas) distributed among groups of randomly rotated fragments (clutter). The time course and accuracy of amoeba detection by humans was measured using a two-alternative forced choice protocol with self-reported confidence and variable image presentation time (20-200 ms), followed by an image mask optimized so as to interrupt visual processing.  Measured psychometric functions were well fit by sigmoidal functions with exponential time constants of 30-91 ms, depending on amoeba complexity. Key aspects of the psychophysical experiments were accounted for by a computational network model, in which simulated responses across retinotopic arrays of orientation-selective elements were modulated by cortical association fields, represented as multiplicative kernels computed from the differences in pairwise edge statistics between target and distractor images. Comparing the experimental and the computational results suggests that each iteration of the lateral interactions takes at least $37.5$ ms of cortical processing time. Our results provide evidence that cortical association fields between orientation selective elements in early visual areas can account for important temporal and task-dependent aspects of the psychometric curves characterizing human contour perception, with the remaining discrepancies postulated to arise from the influence of higher cortical areas.

\section*{Author Summary} 
Current computer vision algorithms reproducing the feed-forward features of the primate visual pathway still fall far behind the capabilities of human subjects in detecting objects in cluttered backgrounds. Here we investigate the possibility that recurrent lateral interactions, long hypothesized to form cortical association fields, can account for the dependence of object detection accuracy on shape complexity and image exposure time.  Cortical association fields are thought to aid object detection by reinforcing global image features that cannot easily be detected by single neurons in feed-forward models.  Our implementation uses the spatial arrangement, relative orientation, and continuity of putative contour elements to compute the lateral contextual support.  We designed synthetic images that allowed us to control object shape and background clutter while eliminating unintentional cues to the presence of an otherwise hidden target. In contrast, real objects can vary uncontrollably in shape, are camouflaged to different degrees by background clutter, and are often associated with non-shape cues, making results using natural image sets difficult to interpret. Our computational model of cortical association fields matches many aspects of the time course and object detection accuracy of human subjects on statistically identical synthetic image sets.  This implies that lateral interactions may selectively reinforce smooth object global boundaries.

\section*{Introduction}
The perception of closed contours is fundamental to object recognition, as revealed by the fact that common object categories can be rapidly detected in black and white line drawings in which all shading and luminance cues have been removed~\cite{Velisavljevic_Elder_2009}.  Cortical association fields, hypothesized to capture spatial correlations between local image features via long-range lateral synaptic interactions, provide a natural substrate for rapid contour perception~\cite{Field_Hayes_Hess_1993_VisRes}.  The link between cortical association fields and contour perception has been investigated through a variety of behavioral, experimental, and theoretical techniques~\cite{Loffler_2008, Hess_Field_1999, Fitzpatrick_2000, Series_Lorenceau_Fregnac_2003}. Psychophysical measurements reveal that the detection of implicit contours, defined by sequences of Gabor-like elements presented against randomly oriented backgrounds, becomes more difficult as the local curvature increases and as the individual Gabor elements are spaced further apart or their alignment is randomly perturbed. This dependence on proximity and relative orientation implies that, in early visual areas, cortical association fields are primarily local and aligned along smooth trajectories~\cite{Field_Hayes_Hess_1993_VisRes, Kovacs_Julesz_1993_PNAS, Pettet_McKee_Grzywacz_1998_VisRes}. In related studies, collinear Gabor patches have been shown to both increase and decrease the contrast detection threshold of a central Gabor patch in a manner that depends on the relative timing, orientation and spatial separation of the flanking elements~\cite{Polat_Sagi_1993, Kapadia_Ito_Gilbert_Westheimer_1995, Polat_Sterkin_Yehezkel_2007}, providing further psychophysical evidence that lateral influences act at early cortical processing stages, although the contribution of collinear facilitation to contour integration remains controversial~\cite{Huang_2007}.
In primary visual cortex (V1), electrophysiological recordings indicate that the responses to optimally oriented and positioned stimuli can be facilitated by flanking stimuli placed outside the classical receptive field center~\cite{Kapadia_Ito_Gilbert_Westheimer_1995, Series_Lorenceau_Fregnac_2003, Bringuier_Chavane_Glaeser_Fregnac_1999, Fitzpatrick_2000}, although these effects have also been ascribed to elongated central receptive fields~\cite{Cavanaugh_2002a,Cavanaugh_2002b} and facilitation has been attributed to increases in baseline activity~\cite{Pooresmaeili_Herrero_Self_Roelfsema_Thiele_2010}. 
Nonetheless, collinear facilitation is consistent with anatomical studies indicating that orientation columns are laterally connected to surrounding columns with similar orientation preference~\cite{Bosking_Fitzpatrick_1997, Gilbert_Weisel_1989, Malach_Harel_Grinvald_1993}.

Because extensive association fields are present in the primary visual cortex~\cite{Bosking_Fitzpatrick_1997, Gilbert_Weisel_1989, Malach_Harel_Grinvald_1993},  lateral interactions may be key to discriminating smooth object boundaries at very fast time scales (of the order of tens of ms), as observed in numerous speed of sight psychophysical experiments~\cite{Hess_Beaudot_Mullen_2001_VisRes, Keysers_Xiao_Foldiak_Perrett_2001, Keysers_Perrett_2002, Bacon-Mace_Macea_Fabre-Thorpe_Thorpe_2005, Velisavljevic_Elder_2009}. Correspondingly, theoretical models have proposed that V1 cortical association fields can be described mathematically on the basis of cocircularity, and that relaxation dynamics based on cocircular association fields can extract global contours by suppressing local variation~\cite{Ben-Shahar_Zucker_2004}. Such models are qualitatively consistent with human judgments as to whether pairs of short line segments belong to the same or separate contours, with human judgments closely following the pairwise statistics of edge segments extracted from natural scenes~\cite{Geisler_Perry_2009}. Further, model cortical association fields, when used to detect implicit contours, can predict key aspects of human psychophysics, particularly the measured dependence on the density of foreground elements relative to background elements~\cite{Pettet_McKee_Grzywacz_1998_VisRes, Mandon_Kreiter_2005}.

In this paper, we extend the above studies by investigating whether model cortical association fields can account not only for dependence of contour perception on intrinsic task difficulty, a relationship that has been previously explored~\cite{Pettet_McKee_Grzywacz_1998_VisRes, Mandon_Kreiter_2005}, but also for the detailed time course of human contour detection, an aspect that has heretofore not been modeled explicitly, although the time-dependent influence of lateral interactions has been determined for several theoretical models~\cite{Ursino_2004,Sterkin_2008}.  
In this work, we employ multiplicative relaxational dynamics to estimate the time course of contour detection from a computational model employing optimized kernels.  Model results are then compared to speed-of-sight measurements from human subjects performing the same contour detection task.  To obtain optimized cortical association fields, we design lateral connectivity patterns using a novel method that exploits the global statistical properties of salient contours relative to background clutter. Our procedure, which can be generalized beyond the present application, can be summarized as follows.  

We begin by generating a large training corpus, divided into target and distractor images, from which we obtain estimates of the pairwise co-occurence probability of oriented edges conditioned on the presence or absence of globally salient contours.  From the difference in these two probability distributions, we construct Object-Distractor Difference (ODD) kernels, which are then convolved with every edge feature to obtain the lateral contextual support at each location and orientation across the entire image.  Edge features that receive substantial contextual support from the surrounding edges are preserved, indicating they are likely to belong to a globally salient contour, whereas edge features receiving minimal contextual support are suppressed, indicating they are more likely to be part of the background clutter.  The lateral contextual support is applied in a multiplicative fashion, so as to prevent the appearance of illusory edges, and the process is iterated several times, mimicking the exchange of information along horizontal connections in the primary visual cortex.  Our method is thus intended to capture the essential computational elements of cortical association fields that are hypothesized to mediate the pop-out of salient contours against cluttered backgrounds. 

To obtain a large number of training images and to better isolate the role of cortical association fields linking low-level visual features, we employ abstract computer-generated shapes consisting of short, smooth contour segments that could either be globally aligned to form wiggly, nearly closed objects ({\em amoebas}), or else randomly rotated to provide a background of locally indistinguishable contour fragments ({\em clutter}).  Amoeba targets lack specific semantic content, presumably reducing the influence of high level cortical areas, such as IT.  However, our computer-generated images would not be expected to eliminate the contribution to contour perception from extrastriate areas~\cite{Bair_2003, Zhang_von_der_Heydt_2010, Schwabe_Obermayer_Angelucci_Bressloff_2006, Angelucci_Levitt_Walton_Bullier_Lund_2002}.  Thus, our model of lateral interactions between orientation-selective neurons is designed to account for just one of several cortical mechanisms that likely contribute to contour perception.

Our amoeba/no-amoeba image set differs from stimuli used in previous psychophysical experiments that employed sequences of Gabor-like elements to represent salient contours against randomly oriented backgrounds~\cite{Field_Hayes_Hess_1993_VisRes, Kovacs_Julesz_1993_PNAS, Pettet_McKee_Grzywacz_1998_VisRes}. An advantage of contours represented by random Gabor fields is that the target and distractor Gabor elements can be distributed at approximately equal densities, thereby precluding the use of local density operators as surrogates for global contour perception~\cite{Field_Hayes_Hess_1993_VisRes}. However, our amoeba/no-amoeba image set is more akin to the natural image sets used in previous speed-of-sight object detection tasks~\cite{Serre_Oliva_Poggio_2007}, particularly with respect to studies employing line drawings derived from natural scenes~\cite{Velisavljevic_Elder_2009}.
Humans can detect closed contours, whether defined by aligned Gabor elements or by continuous line fragments, in less than $200$ ms~\cite{Hess_Beaudot_Mullen_2001_VisRes, Velisavljevic_Elder_2009}, which is shorter than the mean interval between saccadic eye movements~\cite{Martinez-Conde_Macknik_Troncoso_Hubel_2009}, thus mitigating the contribution from visual search.
Like Gabor defined contours, our amoeba/no-amoeba image set implements a pop-out detection task involving readily perceived target shapes whose complexity can be controlled parametrically.

To benchmark the accuracy and the time course of the ODD kernel-based procedure applied to the amoeba/no-amoeba task, we compare our model results to the performance of human subjects on a 2AFC speed-of-sight task in which amoeba/no-amoeba images are presented very briefly side by side, followed by a mask designed to limit the time the visual system is able to process the sensory input~\cite{Hess_Beaudot_Mullen_2001_VisRes, Keysers_Xiao_Foldiak_Perrett_2001, Keysers_Perrett_2002, Bacon-Mace_Macea_Fabre-Thorpe_Thorpe_2005, Velisavljevic_Elder_2009}. Since it takes an estimated $100-300$ ms for activation to spread through the ventral stream of the visual cortex~\cite{Keysers_Xiao_Foldiak_Perrett_2001}, an effective mask presented within this time frame can potentially degrade object detection performance by interfering with the neural processing mechanisms underlying recognition~\cite{Rolls_Tovee_1994, Keysers_Perrett_2002}. By plotting task performance as a function of the stimulus onset asynchrony (SOA)--the interval between image and mask presentation onsets--the resulting psychometric curves are hypothesized to estimate the neural processing time required to reach a given level of classification accuracy.  Amoeba targets of low to moderate complexity were found to reliably pop-out against the background clutter, allowing subjects to achieve near perfect performance at SOAs less than $250$ ms, even when followed by an optimized mask consisting of rotated versions of the target and distractor images~\cite{Hess_Beaudot_Mullen_2001_VisRes}.  Our model cortical association fields were able to account for the dependence of human performance on amoeba complexity as well as for aspects of the time course of contour perception as measured by the improvement in human performance with increasing SOA.  Thus, we present the first network-level computational model to simultaneously account for spatial and temporal aspects of contour perception, as measured in human subjects performing the same contour detection task.  Aspects of the experimental data for which our model fails to account, particularly data showing that human subjects require longer processing times to detect more complex targets, may indicate the possible involvement of extrastriate areas, which may be essential for the perception of more complex shapes.

\begin{figure}[thb]
\begin{center}
\includegraphics[width=0.35\columnwidth]{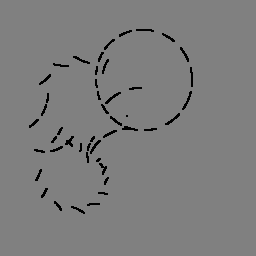}
\hspace{0.03\columnwidth}
\includegraphics[width=0.35\columnwidth]{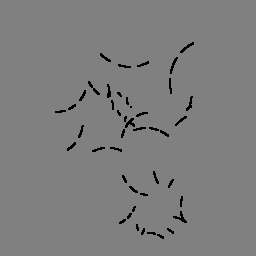}\\
\vspace{0.03\columnwidth}
\includegraphics[width=0.35\columnwidth]{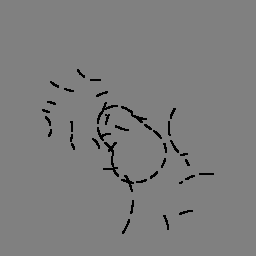}
\hspace{0.03\columnwidth}
\includegraphics[width=0.35\columnwidth]{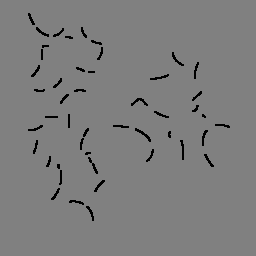}\\
\vspace{0.03\columnwidth}
\includegraphics[width=0.35\columnwidth]{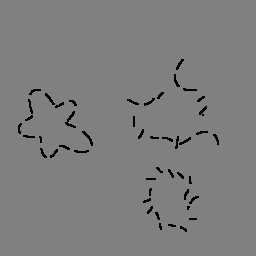}
\hspace{0.03\columnwidth}
\includegraphics[width=0.35\columnwidth]{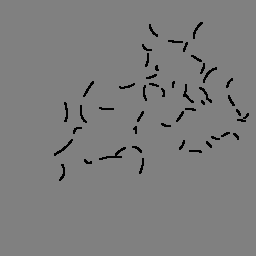}\\
\vspace{0.03\columnwidth}
\includegraphics[width=0.35\columnwidth]{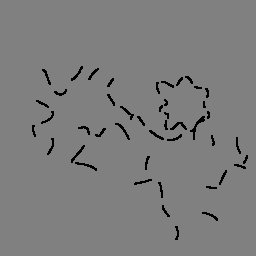}
\hspace{0.03\columnwidth}
\includegraphics[width=0.35\columnwidth]{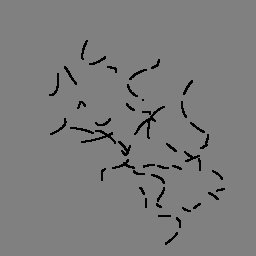}
\end{center}
\caption{ {\bf Examples of targets and distractors from the amoeba/no-amoeba image set for different $K$.} From top to bottom: $K = 2, 4, 6, 8$. Left column: Targets; amoeba complexity increases with increasing numbers of radial frequencies. Clutter was constructed by randomly rotating groups of amoeba contour fragments. Right column: Distractors; only clutter fragments are present.}
\label{fig:amoebas}
\end{figure}

\section*{Results}
To investigate low-level cortical mechanisms for detecting smooth, closed contours presented against cluttered backgrounds with statistically similar low-level features, we designed an amoeba/no-amoeba detection task using a novel set of synthetic images (Figure~\ref{fig:amoebas}). Amoebas are radial frequency patterns~\cite{Wilkinson_Wilson_Habak_1998} constructed via superposition of periodic functions described by a discrete set of radial frequencies around a circle. In addition, we added clutter objects, or distractors, that were locally indistinguishable from targets.  Both targets and distractors were composed of short contour fragments, thus eliminating unambiguous indicators of target presence or absence, such as total line length, the presence of line endpoints, and the existence of short gaps between opposed line segments. To keep the bounding contours smooth, only the lowest $K$ radial frequencies were included in the linear superposition used to construct amoeba targets. To span the maximum range of contour shapes and sizes, the amplitude and phase of each radial frequency component was chosen randomly, under the restriction that the minimum and maximum diameters could not exceed lower and upper limits.  When only $2$ radial frequencies were included in the superposition, the resulting amoebas were very smooth. As more radial frequencies were included, the contours became more complex. Thus, $K$, the number of radial frequencies included in the superposition, provided a control parameter for adjusting target complexity. Figure~\ref{fig:amoebas} shows target and distractor images generated using different values of $K$.

\begin{figure*}[thb]
\begin{center}
\includegraphics[width=1.6\columnwidth]{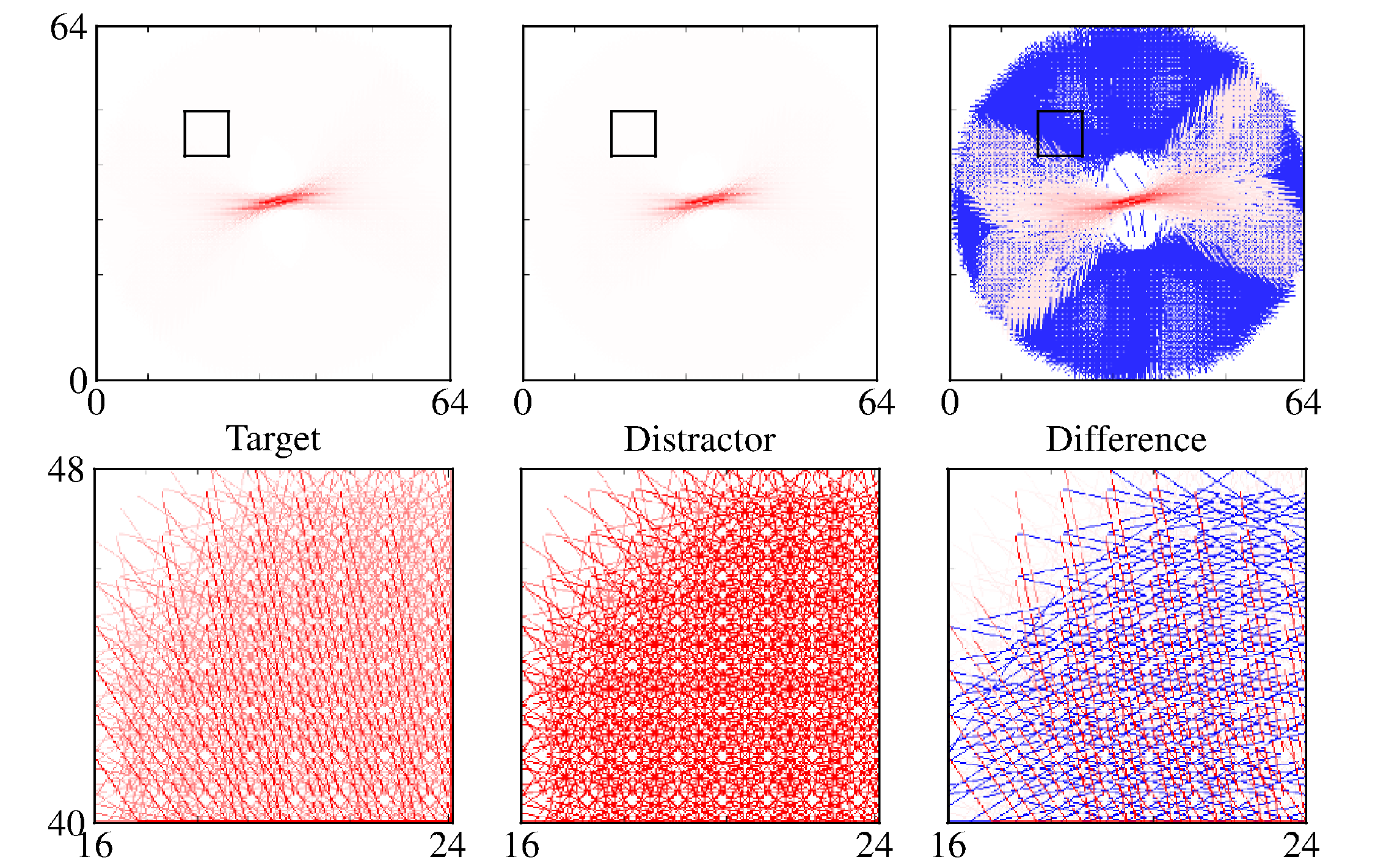}
\end{center}
\caption{ {\bf ODD kernels}. Top Row: For a single short line segment oriented approximately horizontally at the center (not drawn), the co-occurrence-based support of other edges at different relative orientations and spatial locations is depicted. Axes were rotated by ($180^{\circ}/16$) from vertical to mitigate aliasing effects. The color of each edge was set proportional to its co-occurrence-based support. The color scale ranges from blue (negative values) to white (zero) to red (positive values).  Left panel: Co-occurrence statistics compiled from $40,000$ target images. Center panel: Co-occurrence statistics compiled from $40,000$ distractor images. Right panel: ODD kernel, given by the difference in co-occurrence statistics between target and distractor kernels.  Bottom Row: Subfields extracted from the middle of the upper left quadrant (as indicated by black boxes in the top row figures), shown on an expanded scale to better visualize the difference in co-occurrence statistics between target and distractor images. Alignment of edges in target images is mostly cocircular whereas alignment is mostly random in distractor images, accounting for the fine structure in the corresponding section of the ODD kernel.}
\label{fig:fields}
\end{figure*}

Human subjects are able to infer whether a two isolated line segments extracted from a natural scene are from the same or from separate contours using only distance, direction and relative orientation of the two segments as cues~\cite{Geisler_Perry_Super_Gallogly_2001, Geisler_Perry_2009}. The performance of human subjects is well predicted by differences in the empirically calculated co-occurrence statistics of short line segments drawn from either the same or from different contours. To explore the ability of cortical association fields to account for the perception of smooth contours, we developed a network-level computational model of lateral interactions between orientation-selective elements governed by sigmoidal (piecewise linear) input/output synaptic transfer functions. To model lateral interactions, we constructed ``Object-Distractor Difference (ODD) kernels'' for the amoeba/no-amoeba task by computing coactivation statistics for the responses of pairs of orientation-selective filter elements, compiled separately for target and distractor images (Figure~\ref{fig:fields}). 
Because the amoeba/no-amoeba image set was translationally invariant and isotropic, the central filter element may without loss of generality be shifted and rotated to a canonical position and orientation. Thus the canonical ODD kernel was defined relative to filter elements at the origin with orientation $\pi/16$ (to mitigate aliasing effects). Filter elements located away from the origin can be accounted for by a trivial translation. To account for filter elements with different orientations, separate ODD kernels were computed for $8$ orientations then rotated to a common orientation and averaged to produce a canonical ODD kernel. The canonical kernel was then rotated in steps between $0$ and $\pi$ (offset by $\pi/16$) and then interpolated to Cartesian $x-y$ axes by rounding to the nearest integer coordinates.

The resulting ODD kernels were generally consistent with the predictions of cocircular constructions~\cite{Ben-Shahar_Zucker_2004}, except that support was mostly limited to line elements lying along low curvature contours, which follows naturally from the prevalence of low curvatures in our amoeba training set.  

Curiously, the largest differences in the coactivation statistics occur close to the center of the kernel, where targets and distractors are presumably most similar.  However, even at short distances, amoeba segments are still more likely to be aligned than clutter elements.  Moreover, nearby pairs occur much more frequently than more distant pairs, amplifying their contribution to the difference map.  Since, by design, the individual clutter fragments were locally indistinguishable from the target fragments, co-occurrence statistics of oriented fragments were necessary to solve the amoeba/no-amoeba task. The simplest solution, adopted here, was to focus on pairwise co-occurrences. Notably, in some neural preparations, pairwise interactions have been shown to be sufficient to account for a large fraction of all higher-order correlations~\cite{Schneidman_2006,Shlens_Field_Gauthier_Grivich_Petrusca_Sher_Litke_Chichilnisky_2006}.

At the retinal stage, target and distractor images were represented as $256\times 256$ pixel monochromatic, binary line drawings. At the next stage, corresponding to an early cortical processing area such as V1, a set of filters was used to represent $8$ orientations, uniformly-spaced and centered at each pixel, with the axes rotated slightly (by $\pi/16$) to mitigate aliasing artifacts. The bottom-up responses of each orientation-selective element were computed via linear convolution using filters composed of a central excitatory subunit flanked by two inhibitory subunits. Each subunit was an elliptical Gaussian with an aspect ratio of $7:1$, consistent with the aspect ratios of V1 simple cell receptive fields measured experimentally~\cite{Jones_Palmer_1987_JNeurophysiol} and similar to values employed in previously published models of V1 responses~\cite{Troyer_Krukowski_Priebe_Miller_1998_JNeurosci}.  Likewise, we estimate  that each image pixel subtended a visual angle of approximately $0.025^{\circ}$ (see Methods), so that each orientation-selective element in the model subtended a visual angle of approximately $0.2^{\circ}$, consistent with physiological estimates of V1 receptive field sizes at small eccentricities~\cite{Angelucci_Levitt_Walton_Bullier_Lund_2002_JNeurosci}.  All subunits had the same total integrated strength (to within a sign), whose magnitude was adjusted to yield relatively clean representations of the original image in terms of oriented edges. The synaptic transfer function was piecewise-linear with a minimum value of 0.0 and a maximum value of 1.0 and a fixed threshold of 0.5.  A finite threshold and saturation level were essential in order to allow non-supported contour fragments to be suppressed while preventing well-supported fragments from growing without bound.  The precise values used for threshold and saturation were not critical, as responsiveness was controlled independently by adjusting the overall integrated strength of the bottom-up and lateral interaction kernels (see {\em Methods}). 

\begin{figure}[h!]
\begin{center}
\includegraphics[width=0.76\columnwidth]{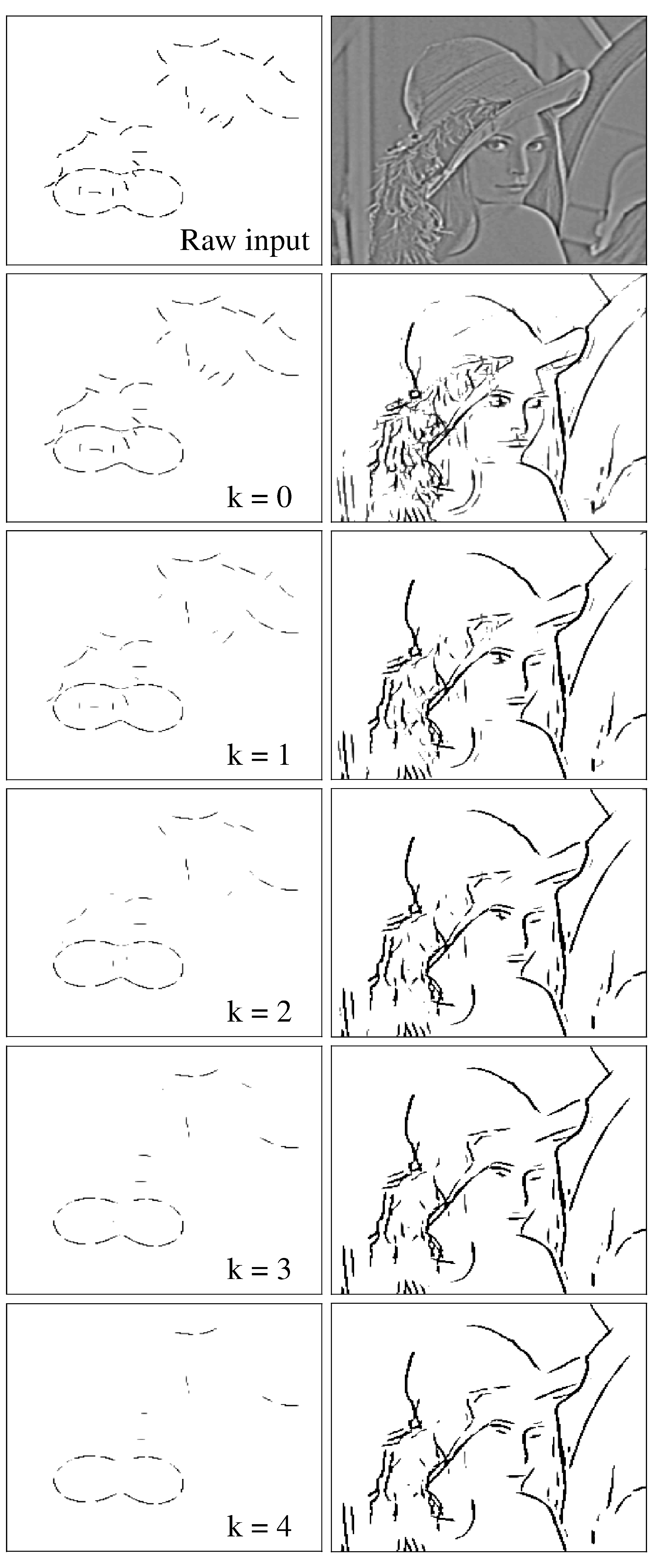}
\end{center}
\caption{ {\bf The effect of lateral interactions on example images.} Left column: black and white amoeba-target image ($K=4$).  Right column: Gray-scale natural image (the standard computer vision test image ``Lena'') after applying a hard Difference of Gaussians (DoG) filter to enhance edges. Top row: Raw retinal input. Second row: Responses of orientation-selective elements before any lateral interactions ($k = 0$). To aid visualization, the activity of the maximally responding orientation-selective element at each pixel location is depicted as a gray-scale intensity. Rows 3-6: Activity after $k = 1, 2, 3, 4$ iterations of the multiplicative ODD kernel, as labeled. For each iteration, activity was multiplied by the local support, computed via linear convolution of the previous output activity with the ODD kernel. Lateral interactions tended to support smooth contours, particularly those arising from amoeba segments, while suppressing clutter or background detail.}
\label{fig:stages}
\end{figure}

Orientation-selective responses were modulated by $4$ successive applications of the multiplicative ODD kernel. Lateral support was first computed via linear convolution of the ODD kernel with the surrounding orientation-selective elements, out to a radius of $32$ pixels.  Given that images were approximately $7^\circ \times 7^\circ$ in extent (see Methods), ODD kernels spanned a total visual angle of approximately $1.75$ degrees, roughly in correspondence with the estimated visuotopic extent of horizontal projections in V1~\cite{Angelucci_Levitt_Walton_Bullier_Lund_2002_JNeurosci}.  The previous activity of each cell was multiplied by the current lateral support, passed through the piecewise-linear synaptic transfer function, and the process repeated for up to $4$ iterations. Contour segments that received insufficient lateral support were thereby suppressed, whereas strongly supported elements were either enhanced or remained maximally activated. When applied to the amoeba/no-amoeba image set, the ODD kernels typically suppressed clutter relative to target segments (Figure~\ref{fig:stages}, left column).

When applied in a similar manner to a natural gray-scale image to which a hard Difference-of-Gaussians (DoG) filter has been applied to maximally enhance local contrast (see Figure~\ref{fig:stages}, right column), ODD-kernels tended to preserve long, smooth lines while suppressing local spatial detail. 
Although ODD kernels were trained on a narrow set of synthetic images, the results exhibit some generalization to natural images due to the overlap between the cocircularity statistics (see Figure~\ref{fig:fields}) of the synthetic image set and those of natural images.

\begin{figure}[h!]
\begin{center}
\includegraphics[width=\columnwidth]{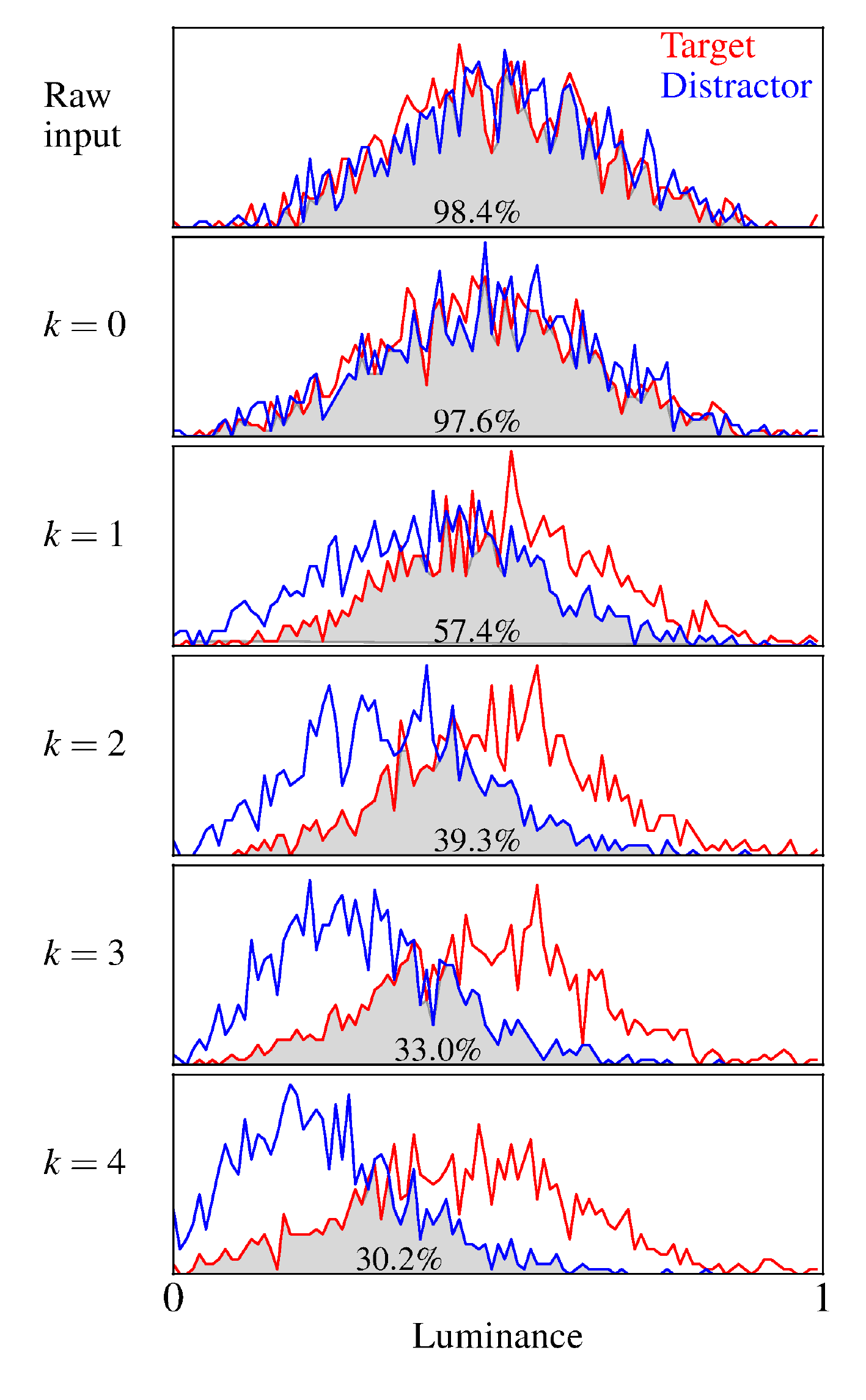}
\end{center}
\caption{ {\bf Histograms of total luminance in target and distractor images as a function of the number of iterations}. Red bins: Total activity histograms for all $1,000$ test target images. Blue bins: Total activity histograms for all $1,000$ test distractor images. The degree that the two distributions overlap is shown as the gray shaded area, which provides a measure of whether total luminance can be used to distinguish targets from distractors.  The percentage in each shaded area shows the approximate lower bound amount of overlap of the two histograms, for comparison.  Top row: Total summed activity over all retinal pixels. Little, if any bias between target and distractor images was evident in the input black and white images as there is nearly complete overlap between the distributions.  Subsequent rows: Total activity histograms summed over all orientation-selective elements. Second row: Bottom-up responses prior to any lateral interactions. Third - sixth rows: Total activity histograms after $1$ - $4$ iterations of the multiplicative ODD kernel, respectively. Total summed activity became progressively more separable with additional iterations, as evinced by a decrease in the overlapping areas.}  
\label{fig:hist}
\end{figure}

To quantify the ability of the model to discriminate between amoeba/no-amoeba target and distractor images, we used the total activation summed over all orientation-selective elements after $k$ iterations of the ODD kernel. A set of $2,000$ target and distractor images was used for testing; test images were generated independently from the training images. Histograms of the total activation show increasing separability between target and distractor images as a function of the number of iterations (Figure~\ref{fig:hist}).  To maximize the range of shapes and sizes spanned by our synthetic targets and distractors, we did not require that the number of ON retinal pixels be constant across images. Rather, the retinal representations of both target and distractor images encompassed a broad range of total activity levels, although the two distributions strongly overlapped and there was no evident bias favoring one or the other. At the next processing stage, prior to any lateral interactions, there was likewise little or no bias evident in the bottom-up responses of the orientation-selective elements.  Each iteration of the multiplicative ODD kernel then caused the distributions of total activity for target and distractor images to become more separable, implying corresponding improvements in discrimination performance on the amoeba/no-amoeba task.

The general principles governing the operation of our model cortical association fields are conceptually straightforward.  ODD kernels, which capture differences in the coactivation statistics of edge segments belonging to amoebas relative to edge segments belonging to the background clutter, are used to determine the lateral contextual support for individual edge segments in an image.  Edge segments receiving sufficiently strong support are preserved, indicating they are likely to be part of an amoeba, whereas edge segments receiving insufficient support are suppressed, indicating they are likely to belong to the background clutter.

\begin{figure}[thb]
\begin{center}
\includegraphics[width=\columnwidth]{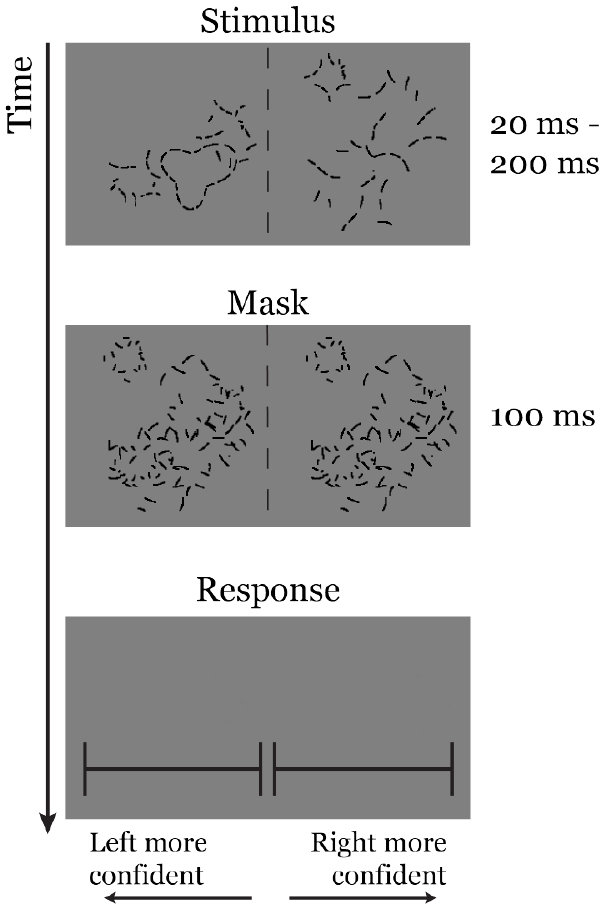}
\end{center}
\caption{{\bf Psychophysical experiment schematic.} The stimulus consisted of one target image and one distractor image (randomly positioned with equal probability on the left or right), presented simultaneously for an SOA between $20$ ms and $200$ ms, followed by an optimized $100$ ms mask generated from randomly rotated groups of target and distractor segments. Subjects indicated which side contained the target object (amoeba) using a computer mouse to click along a horizontal slider bar. Clicking far to the left or right indicated strong confidence that the corresponding side contained the target; clicking close to the center indicated weak confidence. A narrow gap in the center forced subjects to choose between left and right.}
\label{fig:trial}
\end{figure}

To assess the ability of the model cortical association fields to account for the time course of human contour perception, we measured the stimulus presentation time required for human subjects to reach a given level of accuracy on an amoeba/no-amoeba task.  The psychophysical experiment was implemented using a speed-of-sight protocol employing a two-alternative forced choice (2AFC) design, with subjects using a slider bar to indicate which of two images, presented side-by-side, contained an amoeba (Figure~\ref{fig:trial}). The distance the bar was displaced to the left or to the right was used to indicate confidence, see {\em Methods}. To effectively interrupt visual processing at a given SOA, both target and distractor images were replaced by an optimized mask, constructed by combining randomly rotated amoeba and clutter segments~\cite{Hess_Beaudot_Mullen_2001_VisRes}. Our optimized masks were designed to render the amoeba targets virtually invisible in the fused target-mask composite.

\begin{figure*}[thb]
\begin{center}
\includegraphics[width=1.4\columnwidth]{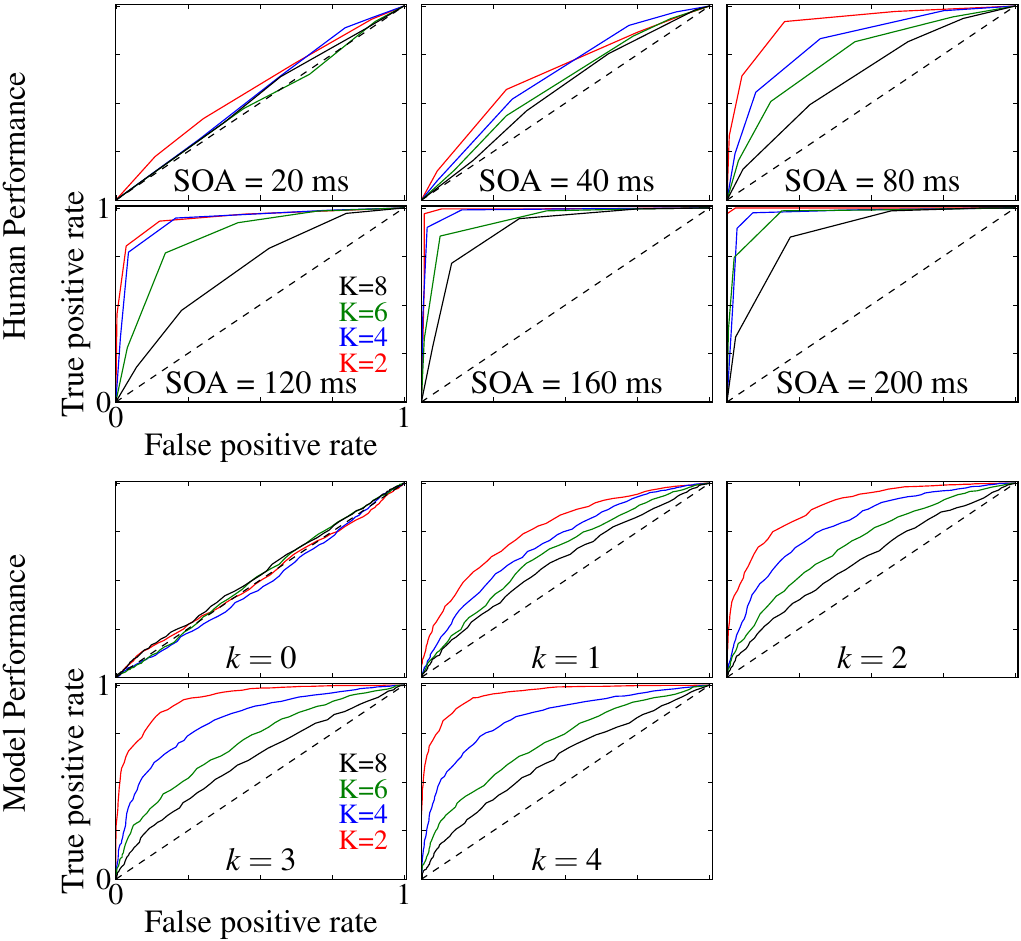}
\end{center}
\caption{{\bf ROC curves comparing human and model performance on the amoeba/no amoeba task.} Top two rows: ROC curves averaged over four different human test subjects using reported confidence (points). The dashed diagonal line in each plot indicates the curve corresponding to chance. Red, blue, green, black correspond to $K=2,4,6,8$, respectively. Bottom two rows: ROC curves for model cortical association fields computed from total activity histograms.}
\label{fig:roc}
\end{figure*}

As a measure of human performance on the amoeba/no-amoeba task, we constructed receiver operating characteristic (ROC) curves~\cite{Azzopardi_Cowey_1997} (Figure~\ref{fig:roc}), using each subject's reported confidence (slider bar location relative to the center position) as a noisy signal for estimating which side, either left or right, contained the target on a given trial.   True positives corresponded to trials on which the subject reported the target was on the left (relative to threshold) and the target was actually on the left (relative to threshold).  False positives corresponded to trials on which the subject reported the target was on the left whereas the target was actually on the right (relative to threshold).  To construct each ROC curve, the confidence scale along the slider bar was divided into 6 discrete threshold values.  For each threshold value, a cumulative proportional true positive rate was calculated by considering only those trials as true positives in which the confidence value was above threshold.  The cumulative proportional false positive rate for each threshold value was calculated similarly.  Each threshold value thus contributed one point on the ROC curve, with true positive rate plotted as the ordinate and the false positive rate as the abscissa.  The complete set of points were connected by straight lines to guide the eye (Figure~\ref{fig:roc}), with a separate ROC curve computed for each combination of SOA and target complexity.  

ROC curves for quantifying the performance of the model on the amoeba/no-amoeba task were computed similarly, using the difference in total luminance between the left and right images as the raw signal for estimating which side contained the target on a given trial.  If the total luminance of the left image was higher than that of the right (relative to threshold), the response of the model would be reported as target on the left.  Ideally, after several iterations of the ODD kernel, no segments would remain in the distractor image and only amoeba segments would remain in the target image; in practice, the total luminance served as a measure of confidence.  Given the much larger number of trials (1000) available for assessing model performance, 100 equally spaced threshold values were used to calculate the corresponding ROC curves.  As with the ROC curves constructed from the confidence values reported by the human subjects, the ROC curves computed from the confidence values reported by the model give the cumulative proportional true positive rate as a function of cumulative proportional false positive rate, with the confidence threshold varied from zero to maximum.  Graphically, the area under the ROC curves is given by the amount of overlap between the total luminance  histograms (see figure~\ref{fig:hist}) for the target and distractor images~\cite{Macmillan_1991}.

ROC curves for human subjects show performance increasingly above chance, indicated by a diagonal line of slope $1$, as a function of both increasing SOA and decreasing target complexity. For amoeba targets of low to moderate complexity, ROC curves obtained from human subjects were well matched to those generated by the model cortical association fields, consistent with the hypothesis that lateral interactions between orientation-selective neurons contribute to human contour perception, at least for simple targets.

\begin{figure*}[thb]
\begin{center}
\includegraphics[width=1.6\columnwidth]{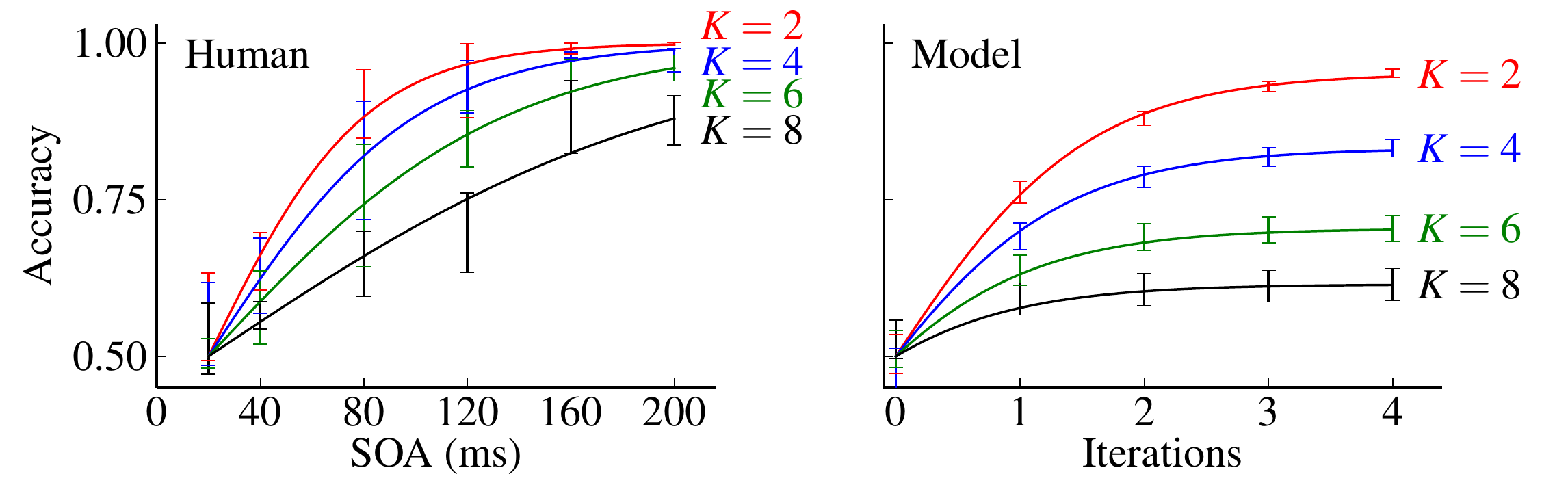}
\end{center}
\caption{ {\bf A comparison of human and model performance on the 2AFC amoeba/no amoeba task.} Left: Average human performance for different SOA in milliseconds. Right: Performance of model cortical association fields for increasing numbers of iterations. Both panels: Accuracy, which is equivalent to area under the ROC curve, (error bars) fitted to single sigmoidal functions (solid lines).  The four curves from top to bottom correspond to $K=2, 4, 6, 8$ radial frequencies.}
\label{fig:fits}
\end{figure*}

The area under the ROC curve (AUC) gives the probability that a randomly chosen target image will be correctly classified relative to a randomly chosen distractor image, and thus provides a threshold-independent assessment of performance on the 2AFC task. Both the average over human subjects and the model cortical association fields exhibited qualitatively similar performance on the 2AFC amoeba/no-amoeba task (Figure~\ref{fig:fits}).  Performance declined as a function of increasing target complexity, both for human subjects, measured at a fixed SOA, and for the model, measured at a fixed number of iterations, implying that $K$ was an effective control parameter for adjusting task difficulty. At $20$ ms SOA, the performance of human subjects was indistinguishable from chance, suggesting that our optimized masks effectively prevented the development of bottom-up cortical responses, even for the simplest targets ($K=2$). Although some studies report that line drawings are processed more rapidly than natural images, with above chance performance being observed at short SOA values~\cite{Velisavljevic_Elder_2009, Mandon_Kreiter_2005}, the fact that performance on the amoeba/no-amoeba task was no better than chance at a $20$ ms SOA implies that our optimized masks effectively interrupted visual processing of the amoeba targets. Since the model used here did not include any account for the time course of bottom-up retinocortical dynamics, we assumed that the performance of human subjects at $20$ ms SOA should be equated to model performance at $0$ iterations (prior to any lateral interactions), a time frame consistent with the distribution of the shortest measured response latencies recorded in primary visual cortex~\cite{Maunsell_1992}.

Overall, average human performance improved as a function of increasing SOA in a manner analogous to the improvement in model performance as a function of the number of iterations of the ODD kernel. This correspondence was especially evident for amoebas of low to moderate complexity ($K\leq 4$ ). For more complex targets, model performance lagged well behind that of human subjects. Studies suggest that low and high radial frequencies are processed by different cortical channels~\cite{Bell_Badcock_Wilson_Wilkinson_2007}. Model performance might have been improved by training a new set of ODD kernels specifically for targets containing $K\geq 6$ radial frequencies, thereby utilizing a hypothetical sub-population of orientation-selective neurons optimized for detecting high-curvature contours. Here, our model was limited to a single multiplicative kernel for detecting all predominately smooth contours. 

To quantify how average human performance on the 2AFC amoeba/no-amoeba task varied with SOA, and to compare with the dependence of model performance on the number of iterations of the ODD kernel, areas under both sets of ROC curves were fit to a monotonically increasing function of the following sigmoidal form:
\begin{equation}
f(t) = \frac{F_{\infty}}{1-(1-2F_{\infty}) e^{-\lambda (t-t_0)}}.
\end{equation}
For human experiments, the parameter $t$ corresponds to the SOA in ms. Since we expect humans to perform close to $100\%$ accuracy for very long SOA, we set $F_{\infty} = 1$. Since humans perform essentially at chance ($50\%$) for $20$ ms SOA, we set $t_0 = 20$ ms.    Thus $\lambda$ was the only free parameter; fits to the average human data were denoted by $\lambda_H$; $\lambda_H$ has units of $1/\text{ms}$. Likewise, model performance was fit to a curve with the same functional form, with $t=k$ measuring the number of iterations; $\lambda_M$ was used to denote curve fits to the model data. However, visual inspection of the model data suggests that its performance saturates at less than $100\%$ accuracy even after an infinite number of iterations, thus we forced the sigmoidal curve fit to the model results to asymptote at the final measured value of AUC: $F_{\infty} = AUC_{k=4}(K)$.  Since the model performs better than chance after only $1$ iteration, we set $t_0=0$.  For both the human experiments and the model performance, the functional form of $f(t)$ ensures that $f(t_{0})=\tfrac{1}{2}$, corresponding to a minimal performance equal to chance.

We find that $1/\lambda_H$ and $1/\lambda_M$ behave quite differently as a function of $K$, the number of radial frequencies used in amoeba generation (Figure~\ref{fig:fits}).  As anticipated for a relaxational process governed by a single kernel, the model data was well described by a single value of $\lambda_M$ (in units of $1/\text{iterations}$), equal to $1.26$.  For the human subjects data, values of $\lambda_H$ increased from $0.034$ to $0.011$ as a function of amoeba complexity, corresponding to lateral processing times of $29.8$ to $90.6$ ms, respectively.  
If human performance depended on only a single set of lateral connections, then, at least in the linear approximation case, we might expect human performance to be well described by a single dominant time constant, representing the dominant eigenmode of the horizontal interactions~\cite{Li_2001, Li_1998}.  Multiple time scales in the human performance case may emerge from any number of  physiological mechanisms not included in the present model, including additional non-linearities in the action of the horizontal connections and/or contributions to contour perception from extrastriate areas.  Our data do not allow us to make a firm distinction between these possibilities.

However, one possible interpretation of the present results is that the perception of simple contours is dominated by relatively fast lateral interactions placed early in the visual processing pathway, thereby accounting for the good fit between the model and experimental results for targets of low to moderate complexity.  Building on this interpretation, we postulate that the perception of more complex contours requires more extensive, and therefore slower, processing mechanisms involving higher cortical areas, thus explaining the discrepancy between model and experimental performance as target complexity increases.  Under the assumption that human perception of simple amoeba targets ($K\leq 4$) depends primarily on recurrent lateral interactions between orientation-selective neurons, we can estimate the time required for each iteration of the multiplicative ODD kernel.  This rate is estimated using the $K=2$ time constants from the fits: $\lambda_{M, K=2}/\lambda_{H} = 37.5$ ms per iteration, a value consistent with estimates of lateral conduction delays within the same cortical area~\cite{Bringuier_Chavane_Glaeser_Fregnac_1999}. 

\begin{figure}[thb]
\begin{center}
\includegraphics[width=\columnwidth]{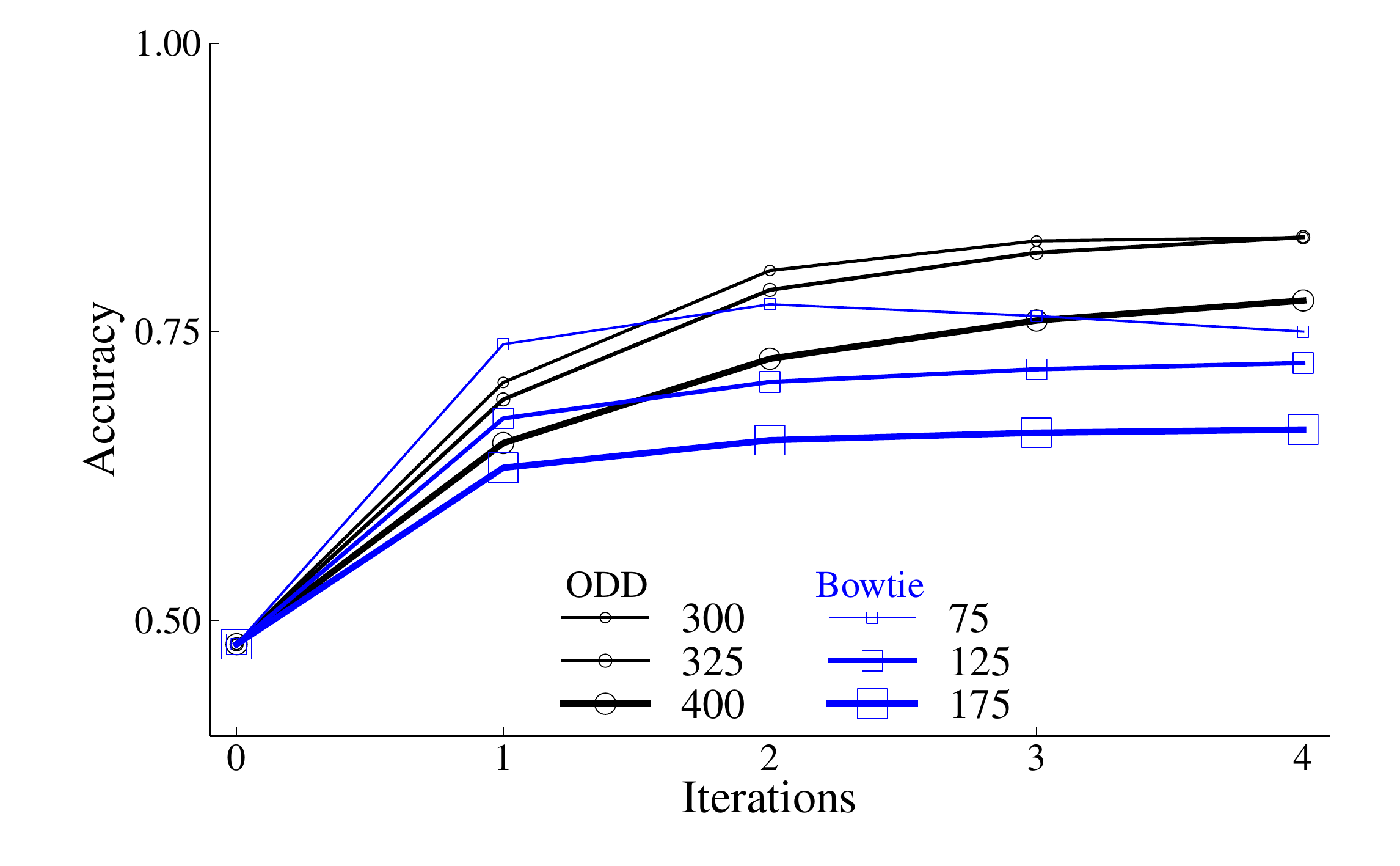}
\end{center}
\caption{ {\bf A comparison of ODD and simpler ``Bowtie'' kernel performance on the on the 2AFC, ${\mathbf K=4}$ 
amoeba/no amoeba task plotted as a function of the number of iterations for a range of different kernel strengths.}  Line width and marker size denote values on kernel strength, which was the main free parameter in the model.  Kernel strength is a dimensionless constant.  Black lines:  ODD kernel performance.  Blue lines:  ``Bowtie'' kernel performance.  Qualitative behavior was similar for both kernels, demonstrating that multiplicative lateral interactions act robustly to reinforce smooth closed contours.}
\label{fig:robust}
\end{figure}

Having shown that the lateral interactions based on multiplicative ODD kernels can account for both spatial and temporal aspects human contour perception, we seek to identify model details that are essential to the performance reported here.  First, we demonstrate that the proposed model is robust and does not require that the magnitude of the ODD kernel be carefully titrated to a precise value.  Model performance on the 2AFC amoeba/no-amoeba task, measured by the area under the ROC curve (AUC) for increasing numbers of iterations $(k=0,1,2,3,4)$, was plotted for different values of the strength of the ODD kernel, given by the total integrated strengths of the equal and opposite target and distractor contributions (Figure~\ref{fig:robust}).  The number of radial frequencies was fixed at $K=4$. Qualitatively similar performance was obtained for ODD kernel strengths ranging from $300$ to $400$.  The ODD kernel used in the present study, whose strength was set to $325$, produced near optimal performance and also exhibited monotonic improvement with increasing numbers of iterations.
That performance was generally insensitive to the value of the main free parameter in the model provides strong evidence for the robustness of the proposed contour detection mechanism based on multiplicative lateral interactions.

A second aspect of the model that merits scrutiny is the detailed structure of the ODD kernels, which were trained using computer-generated images in which the pairwise edge statistics uniquely identifying globally salient contours could be calculated directly.  Previous models of contour perception typically employed much simpler patterns of lateral connectivity, in which excitatory interactions were either collinear or cocircular, and inhibitory interactions were approximately independent of relative orientation~\cite{Li_2001, Li_1998, Mundhenk_2005, Ursino_2004, Pettet_McKee_Grzywacz_1998_VisRes, Ben-Shahar_Zucker_2004}.  To determine if the detailed structure of the ODD kernel was critical to the observed performance, we repeated the amoeba/no-amoeba experiment using a much simpler kernel whose basic form was consistent with a number of previously published models (see Figure~\ref{fig:robust}). Specifically, we used a ``Bowtie''  kernel in which excitatory connections fanned out with an opening angle of $\pi/6$ and the difference in the preferred orientations of the pre- and post-synaptic elements differed by no more than $\pm\pi/6$.  Both excitatory and inhibitory connection strengths fell off in a Gaussian manner, with inhibition strength being insensitive to orientation.  Although the overall accuracy of the Bowtie kernels was lower than that achieved by the ODD kernels, performance on the amoeba/no-amoeba tasks was qualitatively similar, particularly regarding the general monotonic improvement with the number of iterations and the absence of a sensitive dependence on kernel strength.  Thus, we conclude that multiplicative lateral interactions are able to preserve smooth closed contours while suppressing clutter in a manner that is robust to broad changes in model details.

\section*{Discussion}

We have shown that simple models of neural activity in primary visual cortex, enriched with lateral association kernels, reproduce some of the behavioral features regarding the human perception of broken closed contours. Our results agree not only with the measured dependence on contour complexity but also with the temporal dependence of human perception as a function of SOA, suggesting that horizontal connections in V1 may play a non-trivial and global computational role in the perception of closed contours on very fast timescales.

A number of studies relate to the potential contribution of cortical association fields to human contour perception; these encompass a range of anatomical, physiological, psychophysical, and theoretical techniques~\cite{Loffler_2008, Hess_Field_1999, Bringuier_Chavane_Glaeser_Fregnac_1999, Fitzpatrick_2000, Gilbert_Weisel_1989, Malach_Harel_Grinvald_1993, Bosking_Fitzpatrick_1997, Kapadia_Ito_Gilbert_Westheimer_1995, Piech_Gilbert_2006, Polat_Sterkin_Yehezkel_2007, Pooresmaeili_Herrero_Self_Roelfsema_Thiele_2010, Field_Hayes_Hess_1993_VisRes, Kovacs_Julesz_1993_PNAS, Pettet_McKee_Grzywacz_1998_VisRes,Polat_Sagi_1993, Kapadia_Ito_Gilbert_Westheimer_1995}.  In particular, a number of theoretical models have sought to account for human contour perception at the level of biologically-plausible neural circuits~\cite{Grossberg_Mingolla_1985b, Ullman_1992, Pettet_McKee_Grzywacz_1998_VisRes, Yen_Finkel_1998, Garrigues_Olshausen_2007,Mundhenk_2005,Ursino_2004, Sterkin_2008}, with most studies incorporating some form of cortical association field configured to reinforce smoothness~\cite{Ben-Shahar_Zucker_2004}. Although biologically plausible models of cortical association fields have been used to account for the dependence of contour visibility on key parameters controlling task difficulty, such as smoothness, closure, and density of background clutter~\cite{Pettet_McKee_Grzywacz_1998_VisRes}, model cortical association fields have not been directly compared to the time course of human contour perception as a function of contour complexity. Here, we used cortical association fields based on ODD kernels, which were computed from differences in the pairwise coactivation statistics of orientation-selective elements arising from target as opposed to distractor images. While we designed the kernels specifically for the amoeba-clutter disambiguation, we emphasize that the algorithm for the ODD kernel construction is completely general and can be used to improve detection of salient image features in any situation where generative models of targets and distractors are known, or there exists data sets of sufficient size to characterize the contour co-occurrence statistics empirically for both targets and distractors. In our experiments, ODD kernels were able to account for the experimentally observed variations in the saliency of closed contours as a function of parametric complexity and for the time course with which smooth contours are processed by cortical circuits. Crucial for these results was our use of a synthetic target/distractor data set with controllable complexity and the absence of top-down contextual features or local cues that might give away target presence.

Here, we used a semi-supervised training scheme to learn lateral connectivity patterns optimized for performing the amoeba/no-amoeba task.  Necessarily, we sought to model only a subset of the lateral interactions between orientation-selective neurons, namely, those horizontal connections configured to reinforce smooth, closed contours.   We did not attempt to capture the full range of spatial relationships between features extracted at early cortical processing stages~\cite{Ing_Wilson_Geisler_2010_JOV, Ben-Shahar_Zucker_2004}.  Presently, databases containing sufficient numbers of fully {\em annotated} and {\em segmented} natural images needed to reproduce the weeks (or months) of visual experience required to train the full complement of horizontal connections in the primary visual cortex do not exist.  Moreover, the computational resources to exploit such databases, even if they did exist, are highly non-trivial to assemble.  Thus, we focused here on a subset of horizontal connections for which it was possible to construct synthetic surrogate images.  At most, the proposed model represents a subset--and only a subset--of the lateral connections between orientation-selective cortical neurons.  Moreover, even a complete set of such horizontal connections would, at most, represent but a subset of the cortical mechanisms that contribute to the time course and shape-dependence of contour perception.

The supervised training scheme employed here might be related to perceptual learning phenomena, which take place over time scales much shorter than those typically associated with developmental processes~\cite{Yao_Shi_Han_Gao_Dan_2007_NatNeurosci, Hua_Bao_Huang_Wang_Xu_Zhou_Lu_2010_CurrBio,Li_Gilbert_2002_JNeurophysiol}.  It is possible that known physiological mechanisms, such as spike-timing-dependent plasticity (STDP), especially with accounts for realistic conduction delays~\cite{Knoblauch_Sommer_2004}, could mediate a rapid refinement of lateral connections so as to facilitate the perception of amoeba targets.  Moreover, physiological plasticity mechanisms might produce different patterns of connectivity for orientation-selective elements representing points of low as opposed to high local curvature, thereby optimizing lateral interactions for contours of varying complexity.  Here, we made no attempt to customize distinct ODD kernels for detecting contours of varying complexity. Instead, a single ODD kernel was trained using a complete set of images in which different numbers of radial frequency components were equally represented. Although we did not investigate whether, or to what extent, the performance of human subjects improved over the course of the amoeba/no-amoeba experiment, such investigations might shed insight into the role of perceptual learning in the detection of closed contours.      

The question of how lateral connectivity based on ODD kernels might be acquired during development was not addressed explicitly.  In principle, coactivation statistics between pairs of orientation-selective neurons could be accumulated over time in an unsupervised manner by a Hebbian-like learning rule~\cite{Hoyer_Hyvarinen_2002_VisionRes}. Under natural viewing conditions, we expect that contour fragments consistent with smooth, closed boundaries would tend to occur simultaneously, whereas contour fragments inconsistent with object boundaries would tend occur at random temporal delays. Thus, a Hebbian-like learning rule sensitive to temporal correlations, such as certain mathematical forms of STDP-like learning rules~\cite{Song_Abbott_Miller_2000}, might under normal developmental conditions lead to connectivity patterns that reinforce smooth contours.

Of course, human contour perception may have nothing to do with cortical association fields, or lateral interactions may play a subordinate role. Early models showed how spatial filtering could enhance texture-defined contours in the absence of orientation-specific interactions~\cite{Hess_Field_1999} and short-range lateral interactions can accentuate texture-defined boundaries~\cite{Schwabe_Obermayer_Angelucci_Bressloff_2006, Li_2002}. However, psychophysical studies employing implicit contours~\cite{Field_Hayes_Hess_1993_VisRes, Kovacs_Julesz_1993_PNAS, Pettet_McKee_Grzywacz_1998_VisRes}, in which foreground and background elements are present at equal density and which lack explicit texture cues, appear to rule out explanations that omit long-range, orientation-specific interactions. An influential class of biologically-inspired computer vision models achieves a degree of viewpoint-invariant object recognition by constructing feed-forward hierarchies to extract progressively more complex and viewpoint invariant features~\cite{Fukushima_1980,Serre_Oliva_Poggio_2007}. By analogy with such models, scale- and position-independent representations for detecting long, smooth contours could in principle be constructed hierarchically, starting with simple edge detectors and building up progressively longer, more complex curves using a ``bag-of-features'' approach. Presently, there appear to be insufficient data to decide whether human contour perception involves primarily lateral, feed-forward, or even top-down connections~\cite{Angelucci_Levitt_Walton_Bullier_Lund_2002, Gilbert_Sigman_2007, Zhang_von_der_Heydt_2010}. Hypothetically, the cortical association fields used in the present study could have been implemented as a feed-forward architecture, using a hierarchy of orientation-selective neurons to link progressively more widely separated contour fragments. Functionally, there may not exist a clean distinction between lateral, feed-forward and feed-back topologies, with the possibility that all three types of connectivity contribute to human contour perception.

To quantify the temporal dynamics underlying visual processing, we performed speed-of-sight psychophysical experiments that required subjects to detect closed contours (amoebas) spanning a range of shapes, sizes and positions, whose smoothness could be adjusted parametrically by varying the number of radial frequencies (with randomly chosen amplitudes). To better approximate natural viewing conditions, in which target objects usually appear against noisy backgrounds and both foreground and background objects consist of similar low-level visual features, our amoeba/no-amoeba task required amoeba targets to be distinguished from locally indistinguishable open contour fragments (clutter). For amoeba targets consisting of only a few radial frequencies ($K\leq 4$), human subjects were able to perform at close to $100\%$ accuracy after seeing target/distractor image pairs for less than $200$ ms, consistent with a number of studies showing that the recognition of unambiguous targets typically requires $150-250$ ms to reach asymptotic performance~\cite{Rolls_Tovee_1994, Bacon-Mace_Macea_Fabre-Thorpe_Thorpe_2005, Keysers_Perrett_2002}, here likely aided by the high intrinsic saliency of closed shapes relative to open shapes ~\cite{Kovacs_Julesz_1993_PNAS}. Because mean inter-saccade intervals are also in the range of $250$~ms~\cite{Martinez-Conde_Macknik_Troncoso_Hubel_2009}, speed-of-sight studies indicate that unambiguous targets in most natural images can be recognized in a single glance. Similarly, we found that closed contours of low to moderate complexity readily ``pop out'' against background clutter, implying that such radial frequency patterns are processed in parallel, presumably by intrinsic cortical circuitry optimized for automatically extracting smooth, closed contours. As saccadic eye movements were unlikely to play a significant role for such brief presentations, it is unclear to what extent attentional mechanisms are relevant to the speed-of-sight amoeba/no-amoeba task.

Our results further indicate that subjects perform no better than chance at SOAs shorter than approximately $20$ ms. Other studies, however, report above chance performance on unambiguous target detection tasks at similarly short SOA values~\cite{Bacon-Mace_Macea_Fabre-Thorpe_Thorpe_2005, Mandon_Kreiter_2005, Velisavljevic_Elder_2009, Serre_Oliva_Poggio_2007}. The discrepancy may be attributed to the different masks employed. Whereas the above cited studies used masks consisting of either spatially filtered (e.g. $1/f$) noise, distractor images, or scrambled versions of the target image set, we constructed rotation masks that were optimized for each target/distractor image pair~\cite{Hess_Beaudot_Mullen_2001_VisRes}. Our working hypothesis was that an optimized mask should completely obscure the target object in the target-mask composite image; also referred to as pattern masking. The requirement that the mask completely hide the target follows from the assumption that at very short SOA, the target and mask images are likely to be effectively fused due to the finite response time of neurons and receptors in the early visual system~\cite{Schneeweis_Schnapf_1995}. For the amoeba/no-amoeba task, we created optimized masks by rotating the amoeba and clutter fragments with the goal of producing the maximum amount of interference in the responses of orientation-selective cells. Presumably, maximum interference occurs when orientation-selective neurons are presented with randomly rotated contour fragments in rapid succession. Although backward masks can have heterogeneous effects, with performance in some cases showing a $U$-shaped dependence on SOA~\cite{Enns_DiLollo_2000}, for the masks used here performance always increased monotonically with SOA. Empirically, the fact that performance was no better than chance at $20$ ms SOA suggests that our optimized masks were able to effectively interrupt the processing of smooth, closed contours at early cortical processing stages. Indeed, the ability to drive overall performance down to chance at SOA values shorter than $20$  ms could provide an operational criteria for assessing the degree to which a given backward pattern mask is able to effectively interrupt visual processing.

The amoeba/no-amoeba task required the integration of information over length scales spanning viewing angles of approximately $3-4^\circ$, larger than the classical excitatory receptive field size of parafoveal V1 neurons. The amoeba/no-amoeba image set (see Figure~\ref{fig:amoebas}) was configured so that purely local information, such as a few adjoining contour fragments, would not be sufficient to solve the target detection problem. Rather, distinguishing amoebas from clutter required integrating global information across multiple contour fragments. Our results suggest that such global integration can be accomplished via lateral interactions between local, orientation-selective filters. Although the density of target and clutter segments was not precisely equilibrated in our amoeba/no-amoeba image set, the wide range of target sizes and shapes spanned by our image generation algorithm makes it unlikely that the near perfect performance of human subjects at long SOA could have been attained using density cues alone~\cite{Hess_Field_1999}.  Here, lateral inputs were used to modulate the bottom-up responses in a multiplicative fashion, so that our cortical association fields acted primarily as gates that suppressed contour fragments that did not receive sufficiently strong contextual support. By preventing lateral inputs from producing activity unless there was already a strong bottom-up input, a multiplicative non-linearity prevented the activation of contour fragments not present in the original image.

The phenomenon of illusory contours suggests that in some cases contextual effects can produce activity even in the absence of a direct bottom-up response~\cite{Zhang_von_der_Heydt_2010}. The precise form of the multiplicative interaction used here was adopted for algorithmic simplicity rather than for biological realism. We observed that including a small additive contribution from the lateral interactions did not fundamentally affect our conclusions. This suggests that ODD kernels, if implemented more generally, might account for the perception of illusory contours as well. However, a more realistic description of the underlying cellular and synaptic dynamics would likely be necessary to model a relaxation process that includes both additive and multiplicative elements.

Both the model and the psychophysical experiments employed a 2AFC design (see Figure~\ref{fig:trial}) in which the goal was to correctly identify which of a pair of images contained an amoeba target. Since each trial involved a forced choice between two images, the model used a simple classifier that labeled the image with greater total activity as the target. For both human subjects and the model, the number of radial frequencies $K$ proved to be a good control parameter for adjusting task difficulty (see Figure~\ref{fig:fits}). For targets of low to moderate complexity, both model performance (as a function of number of iterations) and human performance (as a function of increasing SOA) monotonically approached nearly perfect asymptotic performance as described by a single sigmoidal function with a characteristic scale, representing either time or number of iterations, that increased with $K$ (see Figure~\ref{fig:fits}). Based on comparison with human performance at different SOA values, each iteration of the ODD kernels was estimated to require approximately $37.6$ ms of cortical processing time, consistent with measured conduction delays between laterally connected cortical neurons~\cite{Bringuier_Chavane_Glaeser_Fregnac_1999}.

Prior to any lateral interactions, the stimulus was projected onto a retinotopic array of orientation-selective filter elements, providing a convenient representation for learning cortical association fields by computing differences in pairwise coactivation statistics between target and distractor images. We found that each iteration of the ODD kernel increased the activity of contour fragments that were part of amoebas compared to the activity of clutter fragments, so that after several iterations the mean overall activity, summed across all orientation-selective filter elements, was higher on average for target images than distractor images (see Figure~\ref{fig:hist}). Even in trials that were incorrectly classified, contour fragments belonging to amoebas were typically still favored relative to background clutter. Because the total number of contour fragments varied from trial to trial, with only the average number of fragments being fixed across the entire image set, our relatively crude criterion for discriminating between target and distractor images sometimes led to classification errors even when amoeba fragments had been partially segmented from the background clutter, simply because the distractor image initially contained more fragments. A more sophisticated classifier might have led to a closer correspondence between model and human performance.  Although performance of the present multiplicative model appeared to saturate after only a few iterations of the ODD kernel (e.g. $k\leq 4$), it is possible that a different implementation might have continued to show improvements after additional iterations.  However, the longer processing time implied by additional iterations suggests that other physiological mechanisms, particularly visual search, would likely come into play.
Granted, there is an apparent mismatch between the fading of clutter elements in the model and the persistence of such elements perceptually in human subjects. To reconcile this apparent mismatch, it has been suggested that the initial perception of brightness might be driven by the initial bottom-up response of the individual orientation-selective feature detectors, whereas persistent responses across these same feature detectors might drive salience~\cite{Sterkin_2008}.

The amoeba/no-amoeba image set was designed to allow for parameterized complexity (in terms of the amount of clutter, number of radial frequencies, etc.) while avoiding reference to exogenous world knowledge. Since the amoeba/no-amoeba image set was machine generated, it was possible to produce a very large number of training images; $40,000$ target and $40,000$ distractor images at $256\times 256$ pixel resolution were used to train ODD kernels in the present study. Many computer vision systems employ standard image classification datasets such as the Caltech $101$~\cite{FeiFei_2004}, which allows for uniform benchmarking and thus facilitates direct comparison between models. Datasets based on natural images, however, suffer from several shortcomings. First, the resolution and number of images are fixed when the set is created. While some man-made datasets, such as MNIST~\cite{LeCun_1998}), consist of tens of thousands of handwritten characters, annotated sets of natural photographs ideal for speed-of-sight experiments are typically limited to a few hundred images. In contrast, humans are exposed to millions of natural scenes during visual development. Biologically motivated models that attempt to replicate human performance might require similar numbers of examples. A second shortcoming of natural image datasets is prevalence of high-level contextual information that utilizes exogenous world knowledge, such as the increased a priori likelihood of finding a car on a road, or an animal in a forest. Exploiting such exogenous world knowledge posses a formidable challenge for existing computational models and, on tasks that employ natural images, may obscure the ability of such models to extract behaviorally meaningful information from low-level visual cues. Third, natural image datasets typically provide limited capability for adjusting intrinsic task difficultly. For example, one widely used dataset~\cite{Serre_Oliva_Poggio_2007} includes photographs of animals at different distances, but only a few discrete distances are annotated and the relationship of target distance to task difficultly is not easily quantified. Here, we illustrated how a synthetic set of images could be used to compare model and human performance in a task with parametric difficulty, potentially validating the use of artificial as opposed to natural images.

The present study addressed the role of cortical association fields in the perception of closed contours, which are presumably important for detecting visual targets based on shape or outline. Although studies show that human subjects can rapidly distinguish between images containing target and non-target object categories using only the line drawings obtained by filtering natural scenes~\cite{Velisavljevic_Elder_2009}, normal experience involves a number of complementary visual cues, such as texture, color, motion and stereopsis. Presumably, cortical association fields also act to reinforce features representing these complementary visual cues as well. Human subjects, for example, can distinguish whether pairs of texture patches were drawn from the same natural object or two different natural objects in a manner that exhibits a similar dependence on pairwise co-occurrence statistics as was found for orientated edges~\cite{Ing_Wilson_Geisler_2010_JOV}. We may speculate that an analysis of coactivation statistics for features selective to a combination of cues such as local orientation, texture, color, motion, and disparity may lead to a more general and more powerful set of kernels capable of fast and effective determination of global object properties, which in turn can play an important role in complex object identification.

\section*{Methods}
\subsection*{Synthetic amoeba/no-amoeba image set}
An {\em amoeba} is a type of radial frequency pattern~\cite{Wilkinson_Wilson_Habak_1998} consisting of a deformed circle in which the radius varies as a function of the polar angle. By choosing the number and relative amplitudes of the different frequency components, the radius can describe an arbitrarily complex shape, exactly analogous to how a Fourier basis can be used to construct an arbitrary waveform on a finite interval. Each radial frequency component was represented by a sinusoidal function defined at $C=1024$ discrete polar angles, spaced uniformly on the interval $[0,2\pi)$. The cutoff radial frequency used in constructing the closed contour provided a control parameter for regulating the complexity of the resulting figure, which ranged from nearly circular, when only the $2$ lowest radial frequencies had non-zero amplitudes, to highly sinusoidal and irregular, when the first $8$ radial frequencies had non-zero amplitudes. All amoeba shapes generated here may be considered smooth, in that local curvature was always bounded.

In detail, the radius of an amoeba at each polar angle was:
\begin{equation}
r(\phi_c)=A_0 + \sum_{n=1}^K A_n \cos \biggl( \frac{2\pi n}{C} + \alpha_n\biggr).
\end{equation}
All amplitudes $A_n$ were initially drawn from normal distributions with $0$ mean and unit variance. All phases $\alpha_n$ were drawn from uniform distributions over the interval $0$ and $\pi/2$. The resulting radial frequency pattern was then linearly rescaled so that the maximum radius, $r_{\max}$, was equal to a random number drawn from a uniform distribution such that $L/4\leq r_{\max}\leq L/2$, where $L$ is the linear size of the square image ($L=256$ pixels), and the minimum radius was given by a second randomly chosen value so that $r_{\max}/4\leq r_{\min}\leq r_{\max}/2$. Uniform pseudo-random numbers were generated by the intrinsic MATLAB $7.0$ function RAND, or its Octave $3.2.3$ equivalent.

To facilitate the construction of locally indistinguishable clutter and model contour occlusion in natural images, amoeba contours were divided into $16$ periodically-spaced fragments by removing short sections whose lengths varied within a specified range. Specifically, the gaps between amoeba fragments varied from $16$ to $32$ in units of discrete polar angle $(2\pi/C)$. Amoeba contours were then broken into fragments by periodically inserting $16$ gaps of variable width ranging from $16$ to $32$, spaced $C/16$ segments apart. Gaps were deleted from the underlying contour, so that the polar angle subtended by each fragment varied in accordance with the changes in preceding gap width. The starting point of the first gap was chosen randomly on the interval ${1, C/G}$, so that over the entire image set the inserted gaps were distributed uniformly around the circle.

To create clutter fragments, an amoeba was first generated using the above procedure. Consecutive amoeba fragments were then grouped, with the number of fragments in each group determined by a Poisson process with a mean value of $2$ and an upper cutoff of $3$. Each group of amoeba fragments was then rotated about its center of mass through random angles on the interval $\pi/8$ to $7\pi/8$. The resulting clutter consisted of the same fragments as the original amoeba but rotated so that collectively the rotated fragments no longer supported the perception of a closed object. Clutter fragments constructed in this manner were thus locally indistinguishable from amoeba fragments. To create clutter in both target and distractor images, several amoebas were first superimposed at random positions and then groups of fragments rotated following the procedure described above. All amoebas contained the same total number of contour fragments (and therefore the same number of gaps) but varied in both maximum diameter and total contour length.

The center of each amoeba was chosen randomly under the restriction that no contour be allowed to cross an image boundary. Specifically, the $x$-coordinate of the amoeba center, $x_0$, was chosen randomly on a restricted interval, $r_{\max} \leq x_0 \leq L-r_{\max}$, and likewise for the $y$-coordinate, $y_0$. When groups of amoeba fragments were randomly rotated to make clutter, portions of a contour belonging to a clutter fragment would occasionally cross an image boundary. In such cases, any out-of-bounds portions of a contour were reflected back into the image region using mirror boundary conditions.

Target images always consisted of $1$ set of amoeba fragments and $2$ sets of clutter fragments. Distractor images consisted of $3$ sets of clutter fragments and thus, averaged over the entire image set, had the same mean luminance and the same variance as the target images. Mask images were constructed following a procedure nearly identical to that used for constructing distractor images, except that mask images consisted of $6$ sets of clutter fragments, obtained by randomly rotating the $6$ original amoeba objects used in constructing the corresponding target and distractor images. All contour fragments were initially represented as a set of points in polar coordinates, corresponding to the radius at each discrete polar angle. Points along the contour were then transformed back to Cartesian coordinates and rounded to the nearest discrete pixel value.  MATLAB scripts for generating the image set used in this study are publicly available at: \texttt{http://petavision.sourceforge.net}.
\subsection*{Ethics statement}
The Los Alamos National Laboratory (LANL) Human Subjects Research Review Board (HSRRB) has reviewed the following experimental protocol and determined that it provides adequate safeguards for protecting the rights and welfare of human subjects involved in the protocol. The protocol was reviewed and approved in compliance with the U.S. Department of Health and Human Services (DHHS) regulations for the Protection of Human Subjects, 45 CFR 46, and in accordance with the LANL Federal Wide Assurance (FWA\#00000362) with the National Institutes of Health/Office for Human Research Protections (NIH/OHRP). The identification number is LANL 08-03 X.  

\subsection*{Human psychophysics}
Human performance was evaluated using two-alternative forced choice (2AFC) psychophysical experiments. There were $5$ subjects, all with normal or corrected-to-normal vision.  One subject only contributed data for a portion of the tested SOAs. Each subject was seated in a dark room, at an approximate distance of $65$ cm from a $19$-inch nominal ($36.2\times 27$ cm actual size) Hitachi $751$ CRT monitor. Images spanned a viewing angle of approximately $7^\circ \times 7^\circ$. The monitor resolution was $1024\times 768$ pixels and the refresh rate was $100$ Hz. The display was driven by a dual-core $3.0$ GHz Mac Pro, with MATLAB $7.6$ running Psychtoolbox~\cite{Brainard_1997}. 

After a short training period to familiarize the subject with the task, one target image and one distractor image were shown side by side, followed by a mask intended to interrupt cognitive processing of the target and distractor images. Two separate sets of experiments were conducted for each subject. In one set, the SOA was chosen randomly from the values ${40, 80, 120}$ ms. For the second set of experiments, the SOA was chosen randomly from the values ${20, 160, 200}$ ms. The duration of the stimulus was always the same as the SOA, and thus both the target and distractor images remained visible until mask onset. The duration of the mask was always $100$ ms. Each subject was shown $1200$ images divided into $10$ blocks of $120$ images, with rest breaks in between blocks (rest break duration was at the discretion of each subject). The pace of the experiment was under the control of the subject, who initiated each trial using the space bar. A small temporal jitter, chosen uniformly between $0$ to $250$ ms, was added to the interval preceding each trial, to prevent entrainment. Task conditions, consisting of variations in both the SOA and the number of radial frequencies $K$, were randomly interleaved such that each condition occurred the same number of times over the course of the entire experiment.

On each trial, subjects indicated which side contained the target, using a mouse-driven slider bar to report confidence (see Figure~\ref{fig:trial}). The reported confidence values were used to construct receiver operating characteristic (ROC) curves, which plot the percentage of true positives (or hits) against the percentage of false positives (or false alarms), with each true/false positive pair obtained by setting a confidence threshold at a different location along the slider bar. A correct response was not necessarily considered a true positive:  to generate one point on the ROC curve, the reported confidence on each trial was measured relative to the current threshold position, which could be to either the left or to the right of center. Thus, a trial might be labeled as incorrect, even though the subject moved the slider bar in the correct direction, as long as the threshold level was not exceeded. Specifically, whenever the reported confidence fell to the left of threshold, the corresponding trial was treated as though the subject reported the target as being to the left, even if the threshold location had been set to the right of center and the confidence bar had actually been slid to the right. Likewise, when the reported confidence fell to the right of the current threshold position, the trial was always treated as if the subject had reported the target to the right, again regardless of how the subject moved the slider bar relative to the center position. By choosing a range of threshold positions, spanning the full range of reported confidence values, a complete ROC curve was obtained. Note that as the threshold was moved closer to the left edge of the slider bar, the percentage of true and false positives both approached minimum values, since only trials with very high reported confidence could contribute to either the true positive or false positive rate (most trials were rejected as either true or false negatives). As the threshold position moved closer to the center of the confidence slider bar, the percentage of true positives increased. Finally, as the threshold was moved closer to the right edge of the slider bar, both the true positive rate and the percentage of false positives approached maximum values. The true positive rate averaged over all false positive rates, or the area under the ROC curve (AUC), was used as an overall measure of subject performance. The AUC is equivalent to the probability that a randomly chosen target image will be correctly classified relative to a randomly chosen distractor image, and thus directly predicts performance on the 2AFC task. Results for each SOA and for each value of $K$ were averaged over $5$ subjects. Error bars denote the standard deviation over the 5 subjects.

\subsection*{Model}
Model cortical association fields were based on differences in the coactivation statistics of orientation-selective filter elements drawn from target and distractor images. Geisler and Perry measured co-occurrence statistics for oriented edges in human segmented natural images~\cite{Geisler_Perry_2009}, and found a close correspondence to human judgments as to whether pairs of short line fragments were drawn from the same or different contours. Thus, we refer to the difference in coactivation statistics between target object and distractor images as Object-Distractor Difference (ODD) kernels. ODD kernels were trained using $40,000$ target and $40,000$ distractor images, each divided into $4$ sets of $10,000$ images each, with each set associated with a different value of $K=2,4,6,8$. The order in which the images were presented had no bearing on the final form of the ODD kernel; that is, there was no temporal component to the training. Training with more images did not substantively improve performance, although small differences were observed in the ODD kernels trained using a smaller number of images ($10,000$ target and $10,000$ distractor images).

Each $256\times 256$ pixel training image activated a regular array of $256\times 256$ retinal elements whose outputs were either $0$ or $1$, depending on whether the corresponding image pixel was ON or OFF, respectively. Each retinal unit activated a local neighborhood of orientation-selective filters, which spanned $8$ angles spaced uniformly between $0$ and $\pi$. To mitigate aliasing effects, the orientation-selective filters were rotated by a small, fixed offset, equal to $\pi/16$, relative to the axis of the training images. All orientation-selective filters were $7\times 7$ pixels in extent and consisted of a central excitatory subunit, represented by an elliptical Gaussian with a standard deviation of $7.0$ in the longest direction and an aspect ratio of $7.0$, flanked by two inhibitory subunits whose shapes were identical to the central excitatory subunit but were offset by $\pm 1.4$ pixels in the direction orthogonal to the preferred axis.

The weight $W_{\theta}(x_{1}-x_{2},y_{1}-y_{2})$, from a retinal element at $(x_{2},y_{2})$ to a filter element at $(x_{1},y_{1})$ with dominant orientation $\theta$, was given by a sum over excitatory and inhibitory subunits:
\begin{align}
&\mat{W}_{\theta}(x_{1}-x_{2},y_{1}-y_{2})=\mat{W}_{\theta}(\vect{r}_{1}-\vect{r}_{2})\nonumber \\
&= A \Biggl\{ \exp{\biggl[\frac{1}{2}\bigl(\vect{r}_{1}-\vect{r}_{2}\bigr)\cdot \mat{R_{\theta}}^{-1} \mat{\sigma}^{-1} \mat{R_{\theta}}\cdot \bigl(\vect{r}_{1}-\vect{r}_{2}\bigr)^T\biggr]}\nn
&- \exp{\biggl[\frac{1}{2}\bigl(\vect{r}_{1}+\vect{f}-\vect{r}_{2}\bigr)\cdot \mat{R_{\theta}}^{-1} \mat{\sigma}^{-1} \mat{R_{\theta}}\cdot \bigl(\vect{r}_{1}+\vect{f}-\vect{r}_{2}\bigr)^T\biggr]}\nn
&- \exp{\biggl[\frac{1}{2}\bigl(\vect{r}_{1}-\vect{f}-\vect{r}_{2}\bigr)\cdot \mat{R_{\theta}}^{-1} \mat{\sigma}^{-1} \mat{R_{\theta}} \cdot \bigl(\vect{r}_{1}-\vect{f}-\vect{r}_{2}\bigr)^T\biggr]}\Biggr\},
\label{eq:weights}
\end{align}
where the position vector is given by $\vect{r}_{i}=[x_{i},y_{i}]$ and the matrix $\mat{\sigma}=\bigl[\begin{smallmatrix}1&0\\0&7\end{smallmatrix}\bigr]$ describes the shape of the elliptical Gaussian subunits for $\theta=0$. In Eq.~\ref{eq:weights}, $\mat{R_{\theta}}$ is a unitary rotation matrix,
\begin{equation}
    \mat{R_{\theta}} = \begin{bmatrix}
\cos \theta &\sin \theta\\
-\sin \theta &\cos \theta
\end{bmatrix},
\end{equation}
and $\vect{f}=[0.0,1.4]$ is a translation vector in the direction orthogonal to the dominant orientation when $\theta=0$. The amplitude $A$ was determined empirically so that the total integrated strength of all excitatory connections made by each retinal unit equaled $20.0$ (and thus the total strength of all inhibitory connections made by each retinal unit equaled $-40.0$). Mirror boundary conditions were used to mitigate edge effects.
The retinal input to each orientation-selective filter element $\mat{s}(x_{1},y_{1},\theta)$ was then given by
\begin{equation}
\mat{s}(x_{1},y_{1},\theta) = \sum_{x_{2}, y_{2}} \mat{W}_{\theta}(x_{1}-x_{2},y_{1}-y_{2}) \mat{I}_{(x_{1},y_{1})}(x_{2},y_{2}),
\end{equation}
where $\mat{I}_{(x_{1},y_{1})}$ is the $7\times 7$ binary input image patch centered on $(x_{1},y_{1})$. The sum is over all pixels $(x_{2},y_{2})$ that are part of this image patch.
The initial output of each orientation-selective filter element $z_{0}(x_{1},y_{1},\theta)$ was obtained by comparing the sum of its excitatory and inhibitory retinal input to a fixed threshold of $0.5$. Values below threshold were set to $0$ whereas values above unity were set to $1.0$. Thus
\begin{equation}
z(x_{1},y_{1},\theta) = g\bigl(\mat{s}(x_{1},y_{1},\theta)\bigr),
\end{equation}
where the function,
\begin{equation}
g(s) = \begin{cases}
0& s < 0.5\\
s& 0.5 \leq s \leq 1.0\\
1& s > 1.0
\end{cases},
\end{equation}
is an element-wise implementation of these thresholds. The responses of all suprathreshold orientation-selective filters contributed to the coactivation statistics, with only the relative distance, direction, and orientation of filter pairs recorded. Because of the threshold condition, only the most active orientation-selective filters contributed to the coactivation statistics.

For every suprathreshold filter element extracted from the $i$-th target image, coactivation statistics were accumulated relative to all surrounding suprathreshold filter elements extracted from the same image. Thus the ODD kernel $G$ is given by
\begin{equation}
G_{\theta-\theta_{0}}^{t_i}\bigl(\rho(x-x_{0},y-y_{0}),\phi_{\theta_{0}}(x-x_{0},y-y_{0})\bigr) = \sum_{x,y} z_{0}(x,y,\theta),
\end{equation}
where the radial distance $\rho$ is a function of the $(x,y)$ coordinates of the two filter elements, the direction $\phi$ is the angle measured relative to $\theta_{0}$, the sum is over all suprathreshold elements within a cutoff radius of $32$, the superscript $t_i$ denotes the $i$-th target image, and the difference in the orientations of the two filter elements $\theta-\theta_{0}$ is taken modulo $\pi$. Because the amoeba/no-amoeba image set was translationally invariant and isotropic, the central filter element may without loss of generality be shifted and rotated to a canonical position and orientation, so that the dependence on ${\rho_{0}, \phi_{0}, \theta_{0}}$ may be omitted. The coactivation statistics for the $i$-th target image can then be written simply as $G_{\theta}^{t_i}(\rho,\phi)$, where $(\rho, \phi)$ gives the distance and direction from the origin to the filter element with orientation $\theta$, given that the filter element at the origin has orientation $\pi/16$. An analogous expression gives the coactivation statistics for the $j$-th distractor image $G_{\theta}^{d_j}(\rho,\phi)$. The ODD kernel $G_{\theta}(\rho,\phi)$ is given by the difference
\begin{equation}
G_{\theta}(\rho,\phi) = W_{+} \sum_{i} G_{\theta}^{t_i}(\rho,\phi) - W_{-} \sum_j G_{\theta}^{d_j}(\rho,\phi),
\end{equation}
where the sums are taken over all target and distractor images and the normalization factors $W_{+}$ and $W_{-}$ are determined empirically so as to yield a total ODD strength of $325$ (see Figure~\ref{fig:robust} and {\em Results}), defined as the sum over all ODD kernel elements arising from either the target or distractor components. By construction, the sum over all ODD kernel elements equals zero, so that the average lateral support for randomly distributed edge fragments would be neutral. Our results did not depend critically on the RMS magnitude of the ODD kernel (see Figure~\ref{fig:robust}). To minimize storage requirements individual connection strengths were stored as unsigned 8-bit integers, so that the results of the present study did not depend on computation of high precision kernels.

As described above, the canonical ODD kernel is defined relative to filter elements at the origin with orientation $\pi/16$. Filter elements located away from the origin can be accounted for by a trivial translation. To account for filter elements with different orientations, separate ODD kernels were computed for all $8$ orientations then rotated to a common orientation and averaged to produce a canonical ODD kernel. The canonical kernel was then rotated in steps between $0$ and $\pi$ (offset by $\pi/16$) and then interpolated to Cartesian $x-y$ axes by rounding to the nearest integer coordinates. Although it has been demonstrated that global contour saliency is enhanced for orientations along the cardinal axes~\cite{Li_Gilbert_2002_JNeurophysiol}, this bias is by construction absent from this model.

ODD kernels were used to compute lateral support for each orientation-selective filter element, via linear convolution. The output of each filter element was then modulated in a multiplicative fashion by the computed lateral support. The procedure was iterated by calculating new values for the lateral support $\vect{s}$, which were again used to modulate filter outputs in a multiplicative fashion:
\begin{equation}
    \vect{s}_{k}(x,y,\theta) =z_{k-1}(x,y,\theta) \sum_{x^\prime , y^\prime , \theta^\prime} G_{\theta}(\rho, \phi_{\theta}) z_{k-1}(x^\prime , y^\prime , \theta^\prime),
\end{equation}
where the subscript $k$ denotes the $k$-th iteration. The same kernel was used for all iterations.  All source code used to train and apply cortical association fields is publicly available at \\\texttt{http://sourceforge.net/projects/petavision/}.

To measure model performance, in each trial $1$ target image and $1$ distractor image were tested as a pair, so as to emulate the 2AFC format of the human experiments. The orientation-selective filter responses to both test images were evaluated after $k={0,1,2,3,4}$ iterations of the ODD kernel. The total activation across all filter elements, $T=\sum_{x,y,\theta'}z_{k}(x,y,\theta')$, was used to compare the two test images. Since the model cortical association fields tended to support contour fragments belonging to amoebas while inhibiting clutter fragments, the image with higher total activation $T$ was assumed to be the target image.  Error bars for the model performance (as shown in Figure~\ref{fig:fits}) were estimated using the standard deviation of a binomial distribution with probability $p$ equal to percent correct and $N$ equal to the number of trials.

\section*{Acknowledgments}
The authors wish to thank Steven Zucker for stimulating discussions that helped initiate this project. 
This work was supported by Los Alamos National Laboratory LDRD program under project 20090006DR; the National Science Foundation, grant ID 0749348; and the DARPA NeoVision2 project.  This publication qualified for unclassified release under DUSA BIOSCI with LA-UR 11-00499.
%
%

\begin{thebibliography}{69}
\expandafter\ifx\csname natexlab\endcsname\relax\def\natexlab#1{#1}\fi
\expandafter\ifx\csname bibnamefont\endcsname\relax
  \def\bibnamefont#1{#1}\fi
\expandafter\ifx\csname bibfnamefont\endcsname\relax
  \def\bibfnamefont#1{#1}\fi
\expandafter\ifx\csname citenamefont\endcsname\relax
  \def\citenamefont#1{#1}\fi
\expandafter\ifx\csname url\endcsname\relax
  \def\url#1{\texttt{#1}}\fi
\expandafter\ifx\csname urlprefix\endcsname\relax\def\urlprefix{URL }\fi
\providecommand{\bibinfo}[2]{#2}
\providecommand{\eprint}[2][]{\url{#2}}

\bibitem[{\citenamefont{Velisavljevi{\'c} and
  Elder}(2009)}]{Velisavljevic_Elder_2009}
\bibinfo{author}{\bibfnamefont{L.}~\bibnamefont{Velisavljevi{\'c}}}
  \bibnamefont{and} \bibinfo{author}{\bibfnamefont{J.~H.} \bibnamefont{Elder}},
  \bibinfo{journal}{J Vision} \textbf{\bibinfo{volume}{9}}
  (\bibinfo{year}{2009}).

\bibitem[{\citenamefont{Field et~al.}(1993)\citenamefont{Field, Hayes, and
  Hess}}]{Field_Hayes_Hess_1993_VisRes}
\bibinfo{author}{\bibfnamefont{D.~J.} \bibnamefont{Field}},
  \bibinfo{author}{\bibfnamefont{A.}~\bibnamefont{Hayes}}, \bibnamefont{and}
  \bibinfo{author}{\bibfnamefont{R.~F.} \bibnamefont{Hess}},
  \bibinfo{journal}{Vision Res} \textbf{\bibinfo{volume}{33}},
  \bibinfo{pages}{173 } (\bibinfo{year}{1993}), ISSN \bibinfo{issn}{0042-6989}.

\bibitem[{\citenamefont{Loffler}(2008)}]{Loffler_2008}
\bibinfo{author}{\bibfnamefont{G.}~\bibnamefont{Loffler}},
  \bibinfo{journal}{Vision Res} \textbf{\bibinfo{volume}{48}},
  \bibinfo{pages}{2106 } (\bibinfo{year}{2008}), ISSN
  \bibinfo{issn}{0042-6989}, \bibinfo{note}{vision Res Reviews}.

\bibitem[{\citenamefont{Hess and Field}(1999)}]{Hess_Field_1999}
\bibinfo{author}{\bibfnamefont{R.}~\bibnamefont{Hess}} \bibnamefont{and}
  \bibinfo{author}{\bibfnamefont{D.}~\bibnamefont{Field}},
  \bibinfo{journal}{Trends Cogn Sci} \textbf{\bibinfo{volume}{3}},
  \bibinfo{pages}{480 } (\bibinfo{year}{1999}), ISSN \bibinfo{issn}{1364-6613}.

\bibitem[{\citenamefont{Fitzpatrick}(2000)}]{Fitzpatrick_2000}
\bibinfo{author}{\bibfnamefont{D.}~\bibnamefont{Fitzpatrick}},
  \bibinfo{journal}{Curr Opin in Neurobiol} \textbf{\bibinfo{volume}{10}},
  \bibinfo{pages}{438 } (\bibinfo{year}{2000}), ISSN \bibinfo{issn}{0959-4388}.

\bibitem[{\citenamefont{Seri{\'e}s et~al.}(2003)\citenamefont{Seri{\'e}s,
  Lorenceau, and Fr{\'e}gnac}}]{Series_Lorenceau_Fregnac_2003}
\bibinfo{author}{\bibfnamefont{P.}~\bibnamefont{Seri{\'e}s}},
  \bibinfo{author}{\bibfnamefont{J.}~\bibnamefont{Lorenceau}},
  \bibnamefont{and}
  \bibinfo{author}{\bibfnamefont{Y.}~\bibnamefont{Fr{\'e}gnac}},
  \bibinfo{journal}{J Physiology-Paris} \textbf{\bibinfo{volume}{97}},
  \bibinfo{pages}{453 } (\bibinfo{year}{2003}), ISSN \bibinfo{issn}{0928-4257},
  \bibinfo{note}{neuroscience and Computation}.

\bibitem[{\citenamefont{Kov{\'a}cs and Julesz}(1993)}]{Kovacs_Julesz_1993_PNAS}
\bibinfo{author}{\bibfnamefont{I.}~\bibnamefont{Kov{\'a}cs}} \bibnamefont{and}
  \bibinfo{author}{\bibfnamefont{B.}~\bibnamefont{Julesz}}, \bibinfo{journal}{P
  Natl Acad Sci USA} \textbf{\bibinfo{volume}{90}}, \bibinfo{pages}{7495}
  (\bibinfo{year}{1993}).

\bibitem[{\citenamefont{Pettet et~al.}(1998)\citenamefont{Pettet, McKee, and
  Grzywacz}}]{Pettet_McKee_Grzywacz_1998_VisRes}
\bibinfo{author}{\bibfnamefont{M.~W.} \bibnamefont{Pettet}},
  \bibinfo{author}{\bibfnamefont{S.~P.} \bibnamefont{McKee}}, \bibnamefont{and}
  \bibinfo{author}{\bibfnamefont{N.~M.} \bibnamefont{Grzywacz}},
  \bibinfo{journal}{Vision Res} \textbf{\bibinfo{volume}{38}},
  \bibinfo{pages}{865 } (\bibinfo{year}{1998}), ISSN \bibinfo{issn}{0042-6989}.

\bibitem[{\citenamefont{Polat and Sagi}(1993)}]{Polat_Sagi_1993}
\bibinfo{author}{\bibfnamefont{U.}~\bibnamefont{Polat}} \bibnamefont{and}
  \bibinfo{author}{\bibfnamefont{D.}~\bibnamefont{Sagi}},
  \bibinfo{journal}{Vision Res} \textbf{\bibinfo{volume}{33}},
  \bibinfo{pages}{993 } (\bibinfo{year}{1993}).

\bibitem[{\citenamefont{Kapadia et~al.}(1995)\citenamefont{Kapadia, Ito,
  Gilbert, and Westheimer}}]{Kapadia_Ito_Gilbert_Westheimer_1995}
\bibinfo{author}{\bibfnamefont{M.~K.} \bibnamefont{Kapadia}},
  \bibinfo{author}{\bibfnamefont{M.}~\bibnamefont{Ito}},
  \bibinfo{author}{\bibfnamefont{C.~D.} \bibnamefont{Gilbert}},
  \bibnamefont{and}
  \bibinfo{author}{\bibfnamefont{G.}~\bibnamefont{Westheimer}},
  \bibinfo{journal}{Neuron} \textbf{\bibinfo{volume}{15}}, \bibinfo{pages}{843}
  (\bibinfo{year}{1995}).

\bibitem[{\citenamefont{Polat et~al.}(2008)\citenamefont{Polat, Terkin, and
  Yehezkel}}]{Polat_Sterkin_Yehezkel_2007}
\bibinfo{author}{\bibfnamefont{U.}~\bibnamefont{Polat}},
  \bibinfo{author}{\bibfnamefont{A.}~\bibnamefont{Terkin}}, \bibnamefont{and}
  \bibinfo{author}{\bibfnamefont{O.}~\bibnamefont{Yehezkel}},
  \bibinfo{journal}{Adv Cogn Psych} \textbf{\bibinfo{volume}{3}},
  \bibinfo{pages}{153} (\bibinfo{year}{2008}).

\bibitem[{\citenamefont{Huang and Hess}(2007)}]{Huang_2007}
\bibinfo{author}{\bibfnamefont{P.-C.} \bibnamefont{Huang}} \bibnamefont{and}
  \bibinfo{author}{\bibfnamefont{R.~F.} \bibnamefont{Hess}},
  \bibinfo{journal}{Vision Res} \textbf{\bibinfo{volume}{47}},
  \bibinfo{pages}{3108} (\bibinfo{year}{2007}).

\bibitem[{\citenamefont{Bringuier et~al.}(1999)\citenamefont{Bringuier,
  Chavane, Glaeser, and Fr{\'e}gnac}}]{Bringuier_Chavane_Glaeser_Fregnac_1999}
\bibinfo{author}{\bibfnamefont{V.}~\bibnamefont{Bringuier}},
  \bibinfo{author}{\bibfnamefont{F.}~\bibnamefont{Chavane}},
  \bibinfo{author}{\bibfnamefont{L.}~\bibnamefont{Glaeser}}, \bibnamefont{and}
  \bibinfo{author}{\bibfnamefont{Y.}~\bibnamefont{Fr{\'e}gnac}},
  \bibinfo{journal}{Science} \textbf{\bibinfo{volume}{283}},
  \bibinfo{pages}{695} (\bibinfo{year}{1999}).

\bibitem[{\citenamefont{Cavanaugh
  et~al.}(2002{\natexlab{a}})\citenamefont{Cavanaugh, Bair, and
  Movshon}}]{Cavanaugh_2002a}
\bibinfo{author}{\bibfnamefont{J.~R.} \bibnamefont{Cavanaugh}},
  \bibinfo{author}{\bibfnamefont{W.}~\bibnamefont{Bair}}, \bibnamefont{and}
  \bibinfo{author}{\bibfnamefont{J.~A.} \bibnamefont{Movshon}},
  \bibinfo{journal}{J. Neurophys.} \textbf{\bibinfo{volume}{88}},
  \bibinfo{pages}{2530} (\bibinfo{year}{2002}{\natexlab{a}}).

\bibitem[{\citenamefont{Cavanaugh
  et~al.}(2002{\natexlab{b}})\citenamefont{Cavanaugh, Bair, and
  Movshon}}]{Cavanaugh_2002b}
\bibinfo{author}{\bibfnamefont{J.~R.} \bibnamefont{Cavanaugh}},
  \bibinfo{author}{\bibfnamefont{W.}~\bibnamefont{Bair}}, \bibnamefont{and}
  \bibinfo{author}{\bibfnamefont{J.~A.} \bibnamefont{Movshon}},
  \bibinfo{journal}{J. Neurophys.} \textbf{\bibinfo{volume}{88}},
  \bibinfo{pages}{2547} (\bibinfo{year}{2002}{\natexlab{b}}).

\bibitem[{\citenamefont{Pooresmaeili et~al.}(2010)\citenamefont{Pooresmaeili,
  Herrero, Self, Roelfsema, and
  Thiele}}]{Pooresmaeili_Herrero_Self_Roelfsema_Thiele_2010}
\bibinfo{author}{\bibfnamefont{A.}~\bibnamefont{Pooresmaeili}},
  \bibinfo{author}{\bibfnamefont{J.~L.} \bibnamefont{Herrero}},
  \bibinfo{author}{\bibfnamefont{M.~W.} \bibnamefont{Self}},
  \bibinfo{author}{\bibfnamefont{P.~R.} \bibnamefont{Roelfsema}},
  \bibnamefont{and} \bibinfo{author}{\bibfnamefont{A.}~\bibnamefont{Thiele}},
  \bibinfo{journal}{J. Neurosci.} \textbf{\bibinfo{volume}{30}},
  \bibinfo{pages}{12745} (\bibinfo{year}{2010}).

\bibitem[{\citenamefont{Bosking et~al.}(1997)\citenamefont{Bosking, Zhang,
  Schofield, and Fitzpatrick}}]{Bosking_Fitzpatrick_1997}
\bibinfo{author}{\bibfnamefont{W.~H.} \bibnamefont{Bosking}},
  \bibinfo{author}{\bibfnamefont{Y.}~\bibnamefont{Zhang}},
  \bibinfo{author}{\bibfnamefont{B.}~\bibnamefont{Schofield}},
  \bibnamefont{and}
  \bibinfo{author}{\bibfnamefont{D.}~\bibnamefont{Fitzpatrick}},
  \bibinfo{journal}{J. Neurosci.} \textbf{\bibinfo{volume}{17}},
  \bibinfo{pages}{2112} (\bibinfo{year}{1997}).

\bibitem[{\citenamefont{Gilbert and Wiesel}(1989)}]{Gilbert_Weisel_1989}
\bibinfo{author}{\bibfnamefont{C.}~\bibnamefont{Gilbert}} \bibnamefont{and}
  \bibinfo{author}{\bibfnamefont{T.}~\bibnamefont{Wiesel}},
  \bibinfo{journal}{J. Neurosci.} \textbf{\bibinfo{volume}{9}},
  \bibinfo{pages}{2432} (\bibinfo{year}{1989}).

\bibitem[{\citenamefont{Malach et~al.}(1993)\citenamefont{Malach, Amir, Harel,
  and Grinvald}}]{Malach_Harel_Grinvald_1993}
\bibinfo{author}{\bibfnamefont{R.}~\bibnamefont{Malach}},
  \bibinfo{author}{\bibfnamefont{Y.}~\bibnamefont{Amir}},
  \bibinfo{author}{\bibfnamefont{M.}~\bibnamefont{Harel}}, \bibnamefont{and}
  \bibinfo{author}{\bibfnamefont{A.}~\bibnamefont{Grinvald}},
  \bibinfo{journal}{P Natl Acad Sci USA} \textbf{\bibinfo{volume}{90}},
  \bibinfo{pages}{10469} (\bibinfo{year}{1993}).

\bibitem[{\citenamefont{Hess et~al.}(2001)\citenamefont{Hess, Beaudot, and
  Mullen}}]{Hess_Beaudot_Mullen_2001_VisRes}
\bibinfo{author}{\bibfnamefont{R.~F.} \bibnamefont{Hess}},
  \bibinfo{author}{\bibfnamefont{W.~H.~A.} \bibnamefont{Beaudot}},
  \bibnamefont{and} \bibinfo{author}{\bibfnamefont{K.~T.}
  \bibnamefont{Mullen}}, \bibinfo{journal}{Vision Res}
  \textbf{\bibinfo{volume}{41}}, \bibinfo{pages}{1023 } (\bibinfo{year}{2001}),
  ISSN \bibinfo{issn}{0042-6989}.

\bibitem[{\citenamefont{Keysers et~al.}(2001)\citenamefont{Keysers, Xiao,
  F{\"o}ldi{\`a}k, and Perrett}}]{Keysers_Xiao_Foldiak_Perrett_2001}
\bibinfo{author}{\bibfnamefont{C.}~\bibnamefont{Keysers}},
  \bibinfo{author}{\bibfnamefont{D.-K.} \bibnamefont{Xiao}},
  \bibinfo{author}{\bibfnamefont{P.}~\bibnamefont{F{\"o}ldi{\`a}k}},
  \bibnamefont{and} \bibinfo{author}{\bibfnamefont{D.~I.}
  \bibnamefont{Perrett}}, \bibinfo{journal}{J Cognitive Neurosci}
  \textbf{\bibinfo{volume}{13}}, \bibinfo{pages}{90} (\bibinfo{year}{2001}).

\bibitem[{\citenamefont{Keysers and Perrett}(2002)}]{Keysers_Perrett_2002}
\bibinfo{author}{\bibfnamefont{C.}~\bibnamefont{Keysers}} \bibnamefont{and}
  \bibinfo{author}{\bibfnamefont{D.~I.} \bibnamefont{Perrett}},
  \bibinfo{journal}{Trends Cogn Sci} \textbf{\bibinfo{volume}{6}},
  \bibinfo{pages}{120 } (\bibinfo{year}{2002}), ISSN \bibinfo{issn}{1364-6613}.

\bibitem[{\citenamefont{Bacon-Mac{\'e}
  et~al.}(2005)\citenamefont{Bacon-Mac{\'e}, Mac{\'e}, Fabre-Thorpe, and
  Thorpe}}]{Bacon-Mace_Macea_Fabre-Thorpe_Thorpe_2005}
\bibinfo{author}{\bibfnamefont{N.}~\bibnamefont{Bacon-Mac{\'e}}},
  \bibinfo{author}{\bibfnamefont{M.~J.-M.} \bibnamefont{Mac{\'e}}},
  \bibinfo{author}{\bibfnamefont{M.}~\bibnamefont{Fabre-Thorpe}},
  \bibnamefont{and} \bibinfo{author}{\bibfnamefont{S.~J.}
  \bibnamefont{Thorpe}}, \bibinfo{journal}{Vision Res}
  \textbf{\bibinfo{volume}{45}}, \bibinfo{pages}{1459 } (\bibinfo{year}{2005}),
  ISSN \bibinfo{issn}{0042-6989}.

\bibitem[{\citenamefont{Ben-Shahar and Zucker}(2004)}]{Ben-Shahar_Zucker_2004}
\bibinfo{author}{\bibfnamefont{O.}~\bibnamefont{Ben-Shahar}} \bibnamefont{and}
  \bibinfo{author}{\bibfnamefont{S.}~\bibnamefont{Zucker}},
  \bibinfo{journal}{Neural Comput} \textbf{\bibinfo{volume}{16}},
  \bibinfo{pages}{445} (\bibinfo{year}{2004}).

\bibitem[{\citenamefont{Geisler and Perry}(2009)}]{Geisler_Perry_2009}
\bibinfo{author}{\bibfnamefont{W.~S.} \bibnamefont{Geisler}} \bibnamefont{and}
  \bibinfo{author}{\bibfnamefont{J.~S.} \bibnamefont{Perry}},
  \bibinfo{journal}{Visual Neurosci} \textbf{\bibinfo{volume}{26}},
  \bibinfo{pages}{109} (\bibinfo{year}{2009}).

\bibitem[{\citenamefont{Mandon and Kreiter}(2005)}]{Mandon_Kreiter_2005}
\bibinfo{author}{\bibfnamefont{S.}~\bibnamefont{Mandon}} \bibnamefont{and}
  \bibinfo{author}{\bibfnamefont{A.~K.} \bibnamefont{Kreiter}},
  \bibinfo{journal}{Vision Res} \textbf{\bibinfo{volume}{45}},
  \bibinfo{pages}{291 } (\bibinfo{year}{2005}), ISSN \bibinfo{issn}{0042-6989}.

\bibitem[{\citenamefont{Ursino and Cara}(2004)}]{Ursino_2004}
\bibinfo{author}{\bibfnamefont{M.}~\bibnamefont{Ursino}} \bibnamefont{and}
  \bibinfo{author}{\bibfnamefont{G.~E.~L.} \bibnamefont{Cara}},
  \bibinfo{journal}{Neural Networks} \textbf{\bibinfo{volume}{17}},
  \bibinfo{pages}{719} (\bibinfo{year}{2004}).

\bibitem[{\citenamefont{Sterkin et~al.}(2008)\citenamefont{Sterkin, Sterkin,
  and Polat}}]{Sterkin_2008}
\bibinfo{author}{\bibfnamefont{A.}~\bibnamefont{Sterkin}},
  \bibinfo{author}{\bibfnamefont{A.}~\bibnamefont{Sterkin}}, \bibnamefont{and}
  \bibinfo{author}{\bibfnamefont{U.}~\bibnamefont{Polat}}, \bibinfo{journal}{J.
  Vis.} \textbf{\bibinfo{volume}{8}}, \bibinfo{pages}{1}
  (\bibinfo{year}{2008}).

\bibitem[{\citenamefont{Bair et~al.}(2003)\citenamefont{Bair, Cavanaugh, and
  Movshon}}]{Bair_2003}
\bibinfo{author}{\bibfnamefont{W.}~\bibnamefont{Bair}},
  \bibinfo{author}{\bibfnamefont{J.~R.} \bibnamefont{Cavanaugh}},
  \bibnamefont{and} \bibinfo{author}{\bibfnamefont{J.~A.}
  \bibnamefont{Movshon}}, \bibinfo{journal}{J Neurosci}
  \textbf{\bibinfo{volume}{23}}, \bibinfo{pages}{7690} (\bibinfo{year}{2003}).

\bibitem[{\citenamefont{Zhang and von~der
  Heydt}(2010)}]{Zhang_von_der_Heydt_2010}
\bibinfo{author}{\bibfnamefont{N.~R.} \bibnamefont{Zhang}} \bibnamefont{and}
  \bibinfo{author}{\bibfnamefont{R.}~\bibnamefont{von~der Heydt}},
  \bibinfo{journal}{J. Neurosci.} \textbf{\bibinfo{volume}{30}},
  \bibinfo{pages}{6482} (\bibinfo{year}{2010}).

\bibitem[{\citenamefont{Schwabe et~al.}(2006)\citenamefont{Schwabe, Obermayer,
  Angelucci, and Bressloff}}]{Schwabe_Obermayer_Angelucci_Bressloff_2006}
\bibinfo{author}{\bibfnamefont{L.}~\bibnamefont{Schwabe}},
  \bibinfo{author}{\bibfnamefont{K.}~\bibnamefont{Obermayer}},
  \bibinfo{author}{\bibfnamefont{A.}~\bibnamefont{Angelucci}},
  \bibnamefont{and} \bibinfo{author}{\bibfnamefont{P.~C.}
  \bibnamefont{Bressloff}}, \bibinfo{journal}{J. Neurosci.}
  \textbf{\bibinfo{volume}{26}}, \bibinfo{pages}{9117} (\bibinfo{year}{2006}).

\bibitem[{\citenamefont{Angelucci
  et~al.}(2002{\natexlab{a}})\citenamefont{Angelucci, Levitt, Walton, Hupe,
  Bullier, and Lund}}]{Angelucci_Levitt_Walton_Bullier_Lund_2002}
\bibinfo{author}{\bibfnamefont{A.}~\bibnamefont{Angelucci}},
  \bibinfo{author}{\bibfnamefont{J.~B.} \bibnamefont{Levitt}},
  \bibinfo{author}{\bibfnamefont{E.~J.~S.} \bibnamefont{Walton}},
  \bibinfo{author}{\bibfnamefont{J.-M.} \bibnamefont{Hupe}},
  \bibinfo{author}{\bibfnamefont{J.}~\bibnamefont{Bullier}}, \bibnamefont{and}
  \bibinfo{author}{\bibfnamefont{J.~S.} \bibnamefont{Lund}},
  \bibinfo{journal}{J. Neurosci.} \textbf{\bibinfo{volume}{22}},
  \bibinfo{pages}{8633} (\bibinfo{year}{2002}{\natexlab{a}}).

\bibitem[{\citenamefont{Serre et~al.}(2007)\citenamefont{Serre, Oliva, and
  Poggio}}]{Serre_Oliva_Poggio_2007}
\bibinfo{author}{\bibfnamefont{T.}~\bibnamefont{Serre}},
  \bibinfo{author}{\bibfnamefont{A.}~\bibnamefont{Oliva}}, \bibnamefont{and}
  \bibinfo{author}{\bibfnamefont{T.}~\bibnamefont{Poggio}}, \bibinfo{journal}{P
  Natl Acad Sci USA} \textbf{\bibinfo{volume}{104}}, \bibinfo{pages}{6424}
  (\bibinfo{year}{2007}).

\bibitem[{\citenamefont{Martinez-Conde
  et~al.}(2009)\citenamefont{Martinez-Conde, Macknik, Troncoso, and
  Hubel}}]{Martinez-Conde_Macknik_Troncoso_Hubel_2009}
\bibinfo{author}{\bibfnamefont{S.}~\bibnamefont{Martinez-Conde}},
  \bibinfo{author}{\bibfnamefont{S.~L.} \bibnamefont{Macknik}},
  \bibinfo{author}{\bibfnamefont{X.~G.} \bibnamefont{Troncoso}},
  \bibnamefont{and} \bibinfo{author}{\bibfnamefont{D.~H.} \bibnamefont{Hubel}},
  \bibinfo{journal}{Trends Neurosci} \textbf{\bibinfo{volume}{32}},
  \bibinfo{pages}{463 } (\bibinfo{year}{2009}), ISSN \bibinfo{issn}{0166-2236}.

\bibitem[{\citenamefont{Rolls and Tovee}(1994)}]{Rolls_Tovee_1994}
\bibinfo{author}{\bibfnamefont{E.~T.} \bibnamefont{Rolls}} \bibnamefont{and}
  \bibinfo{author}{\bibfnamefont{M.~J.} \bibnamefont{Tovee}},
  \bibinfo{journal}{P Roy Soc Lond B Bio} \textbf{\bibinfo{volume}{257}},
  \bibinfo{pages}{9} (\bibinfo{year}{1994}).

\bibitem[{\citenamefont{Wilkinson et~al.}(1998)\citenamefont{Wilkinson, Wilson,
  and Habak}}]{Wilkinson_Wilson_Habak_1998}
\bibinfo{author}{\bibfnamefont{F.}~\bibnamefont{Wilkinson}},
  \bibinfo{author}{\bibfnamefont{H.~R.} \bibnamefont{Wilson}},
  \bibnamefont{and} \bibinfo{author}{\bibfnamefont{C.}~\bibnamefont{Habak}},
  \bibinfo{journal}{Vision Res} \textbf{\bibinfo{volume}{38}},
  \bibinfo{pages}{3555 } (\bibinfo{year}{1998}), ISSN
  \bibinfo{issn}{0042-6989}.

\bibitem[{\citenamefont{Geisler et~al.}(2001)\citenamefont{Geisler, Perry,
  Super, and Gallogly}}]{Geisler_Perry_Super_Gallogly_2001}
\bibinfo{author}{\bibfnamefont{W.~S.} \bibnamefont{Geisler}},
  \bibinfo{author}{\bibfnamefont{J.~S.} \bibnamefont{Perry}},
  \bibinfo{author}{\bibfnamefont{B.~J.} \bibnamefont{Super}}, \bibnamefont{and}
  \bibinfo{author}{\bibfnamefont{D.~P.} \bibnamefont{Gallogly}},
  \bibinfo{journal}{Vision Res} \textbf{\bibinfo{volume}{41}},
  \bibinfo{pages}{711 } (\bibinfo{year}{2001}), ISSN \bibinfo{issn}{0042-6989}.

\bibitem[{\citenamefont{Schneidman et~al.}(2006)\citenamefont{Schneidman,
  {Berry II}, Segev, and Bialek}}]{Schneidman_2006}
\bibinfo{author}{\bibfnamefont{E.}~\bibnamefont{Schneidman}},
  \bibinfo{author}{\bibfnamefont{M.~J.} \bibnamefont{{Berry II}}},
  \bibinfo{author}{\bibfnamefont{R.}~\bibnamefont{Segev}}, \bibnamefont{and}
  \bibinfo{author}{\bibfnamefont{W.}~\bibnamefont{Bialek}},
  \bibinfo{journal}{Nature} \textbf{\bibinfo{volume}{440}},
  \bibinfo{pages}{1007} (\bibinfo{year}{2006}).

\bibitem[{\citenamefont{Shlens et~al.}(2006)\citenamefont{Shlens, Field,
  Gauthier, Grivich, Petrusca, Sher, Litke, and
  Chichilnisky}}]{Shlens_Field_Gauthier_Grivich_Petrusca_Sher_Litke_Chichilnisky_2006}
\bibinfo{author}{\bibfnamefont{J.}~\bibnamefont{Shlens}},
  \bibinfo{author}{\bibfnamefont{G.~D.} \bibnamefont{Field}},
  \bibinfo{author}{\bibfnamefont{J.~L.} \bibnamefont{Gauthier}},
  \bibinfo{author}{\bibfnamefont{M.~I.} \bibnamefont{Grivich}},
  \bibinfo{author}{\bibfnamefont{D.}~\bibnamefont{Petrusca}},
  \bibinfo{author}{\bibfnamefont{A.}~\bibnamefont{Sher}},
  \bibinfo{author}{\bibfnamefont{A.~M.} \bibnamefont{Litke}}, \bibnamefont{and}
  \bibinfo{author}{\bibfnamefont{E.~J.} \bibnamefont{Chichilnisky}},
  \bibinfo{journal}{J. Neurosci.} \textbf{\bibinfo{volume}{26}},
  \bibinfo{pages}{8254} (\bibinfo{year}{2006}).

\bibitem[{\citenamefont{Jones and
  Palmer}(1987)}]{Jones_Palmer_1987_JNeurophysiol}
\bibinfo{author}{\bibfnamefont{J.~P.} \bibnamefont{Jones}} \bibnamefont{and}
  \bibinfo{author}{\bibfnamefont{L.~A.} \bibnamefont{Palmer}},
  \bibinfo{journal}{J. Neurophys.} \textbf{\bibinfo{volume}{58}},
  \bibinfo{pages}{1233} (\bibinfo{year}{1987}).

\bibitem[{\citenamefont{Troyer et~al.}(1998)\citenamefont{Troyer, Krukowski,
  Priebe, and Miller}}]{Troyer_Krukowski_Priebe_Miller_1998_JNeurosci}
\bibinfo{author}{\bibfnamefont{T.~W.} \bibnamefont{Troyer}},
  \bibinfo{author}{\bibfnamefont{A.~E.} \bibnamefont{Krukowski}},
  \bibinfo{author}{\bibfnamefont{N.~J.} \bibnamefont{Priebe}},
  \bibnamefont{and} \bibinfo{author}{\bibfnamefont{K.~D.}
  \bibnamefont{Miller}}, \bibinfo{journal}{J Neurosci}
  \textbf{\bibinfo{volume}{18}}, \bibinfo{pages}{5908} (\bibinfo{year}{1998}).

\bibitem[{\citenamefont{Angelucci
  et~al.}(2002{\natexlab{b}})\citenamefont{Angelucci, Levitt, Walton, Hup{\'e},
  Bullier, and Lund}}]{Angelucci_Levitt_Walton_Bullier_Lund_2002_JNeurosci}
\bibinfo{author}{\bibfnamefont{A.}~\bibnamefont{Angelucci}},
  \bibinfo{author}{\bibfnamefont{J.~B.} \bibnamefont{Levitt}},
  \bibinfo{author}{\bibfnamefont{E.~J.~S.} \bibnamefont{Walton}},
  \bibinfo{author}{\bibfnamefont{J.-M.} \bibnamefont{Hup{\'e}}},
  \bibinfo{author}{\bibfnamefont{J.}~\bibnamefont{Bullier}}, \bibnamefont{and}
  \bibinfo{author}{\bibfnamefont{J.~S.} \bibnamefont{Lund}},
  \bibinfo{journal}{J Neurosci} \textbf{\bibinfo{volume}{22}},
  \bibinfo{pages}{8633} (\bibinfo{year}{2002}{\natexlab{b}}).

\bibitem[{\citenamefont{Azzopardi and Cowey}(1997)}]{Azzopardi_Cowey_1997}
\bibinfo{author}{\bibfnamefont{P.}~\bibnamefont{Azzopardi}} \bibnamefont{and}
  \bibinfo{author}{\bibfnamefont{A.}~\bibnamefont{Cowey}}, \bibinfo{journal}{P
  Natl Acad Sci USA} \textbf{\bibinfo{volume}{94}}, \bibinfo{pages}{14190}
  (\bibinfo{year}{1997}).

\bibitem[{\citenamefont{Macmillan and Creelman}(1991)}]{Macmillan_1991}
\bibinfo{author}{\bibfnamefont{N.~A.} \bibnamefont{Macmillan}}
  \bibnamefont{and} \bibinfo{author}{\bibfnamefont{C.~D.}
  \bibnamefont{Creelman}}, \emph{\bibinfo{title}{Detection theory: a user's
  guide}} (\bibinfo{publisher}{CUP Archive}, \bibinfo{address}{Cambridge},
  \bibinfo{year}{1991}).

\bibitem[{\citenamefont{Maunsell and Gibson}(1992)}]{Maunsell_1992}
\bibinfo{author}{\bibfnamefont{J.~H.~R.} \bibnamefont{Maunsell}}
  \bibnamefont{and} \bibinfo{author}{\bibfnamefont{J.~R.}
  \bibnamefont{Gibson}}, \bibinfo{journal}{J. Neurophys.}
  \textbf{\bibinfo{volume}{68}}, \bibinfo{pages}{1332} (\bibinfo{year}{1992}).

\bibitem[{\citenamefont{Bell et~al.}(2007)\citenamefont{Bell, Badcock, Wilson,
  and Wilkinson}}]{Bell_Badcock_Wilson_Wilkinson_2007}
\bibinfo{author}{\bibfnamefont{J.}~\bibnamefont{Bell}},
  \bibinfo{author}{\bibfnamefont{D.~R.} \bibnamefont{Badcock}},
  \bibinfo{author}{\bibfnamefont{H.}~\bibnamefont{Wilson}}, \bibnamefont{and}
  \bibinfo{author}{\bibfnamefont{F.}~\bibnamefont{Wilkinson}},
  \bibinfo{journal}{Vision Res} \textbf{\bibinfo{volume}{47}},
  \bibinfo{pages}{1518 } (\bibinfo{year}{2007}), ISSN
  \bibinfo{issn}{0042-6989}.

\bibitem[{\citenamefont{Li}(2001)}]{Li_2001}
\bibinfo{author}{\bibfnamefont{Z.}~\bibnamefont{Li}}, \bibinfo{journal}{Neural
  Comput} \textbf{\bibinfo{volume}{13}}, \bibinfo{pages}{1749}
  (\bibinfo{year}{2001}).

\bibitem[{\citenamefont{Li}(1998)}]{Li_1998}
\bibinfo{author}{\bibfnamefont{Z.}~\bibnamefont{Li}}, \bibinfo{journal}{Neural
  Comput} \textbf{\bibinfo{volume}{10}}, \bibinfo{pages}{903}
  (\bibinfo{year}{1998}).

\bibitem[{\citenamefont{Mundhenk and Itti}(2005)}]{Mundhenk_2005}
\bibinfo{author}{\bibfnamefont{T.~N.} \bibnamefont{Mundhenk}} \bibnamefont{and}
  \bibinfo{author}{\bibfnamefont{L.}~\bibnamefont{Itti}},
  \bibinfo{journal}{Biol Cybern} \textbf{\bibinfo{volume}{93}},
  \bibinfo{pages}{188} (\bibinfo{year}{2005}).

\bibitem[{\citenamefont{Li et~al.}(2006)\citenamefont{Li, Pi{\"e}ch, and
  Gilbert}}]{Piech_Gilbert_2006}
\bibinfo{author}{\bibfnamefont{W.}~\bibnamefont{Li}},
  \bibinfo{author}{\bibfnamefont{V.}~\bibnamefont{Pi{\"e}ch}},
  \bibnamefont{and} \bibinfo{author}{\bibfnamefont{C.~D.}
  \bibnamefont{Gilbert}}, \bibinfo{journal}{Neuron}
  \textbf{\bibinfo{volume}{50}}, \bibinfo{pages}{951} (\bibinfo{year}{2006}).

\bibitem[{\citenamefont{Grossberg and
  Mingolla}(1985)}]{Grossberg_Mingolla_1985b}
\bibinfo{author}{\bibfnamefont{S.}~\bibnamefont{Grossberg}} \bibnamefont{and}
  \bibinfo{author}{\bibfnamefont{E.}~\bibnamefont{Mingolla}},
  \bibinfo{journal}{Percept Psychophys} \textbf{\bibinfo{volume}{38}},
  \bibinfo{pages}{141} (\bibinfo{year}{1985}).

\bibitem[{\citenamefont{Ullman et~al.}(1992)\citenamefont{Ullman, Gregory, and
  Atkinson}}]{Ullman_1992}
\bibinfo{author}{\bibfnamefont{S.}~\bibnamefont{Ullman}},
  \bibinfo{author}{\bibfnamefont{R.~L.} \bibnamefont{Gregory}},
  \bibnamefont{and} \bibinfo{author}{\bibfnamefont{J.}~\bibnamefont{Atkinson}},
  \bibinfo{journal}{Philos T R Soc Lon B} \textbf{\bibinfo{volume}{337}},
  \bibinfo{pages}{371} (\bibinfo{year}{1992}).

\bibitem[{\citenamefont{Yen and Finkel}(1998)}]{Yen_Finkel_1998}
\bibinfo{author}{\bibfnamefont{S.-C.} \bibnamefont{Yen}} \bibnamefont{and}
  \bibinfo{author}{\bibfnamefont{L.~H.} \bibnamefont{Finkel}},
  \bibinfo{journal}{Vision Res} \textbf{\bibinfo{volume}{38}},
  \bibinfo{pages}{719 } (\bibinfo{year}{1998}), ISSN \bibinfo{issn}{0042-6989}.

\bibitem[{\citenamefont{Garrigues and
  Olshausen}(2007)}]{Garrigues_Olshausen_2007}
\bibinfo{author}{\bibfnamefont{P.~J.} \bibnamefont{Garrigues}}
  \bibnamefont{and} \bibinfo{author}{\bibfnamefont{B.~A.}
  \bibnamefont{Olshausen}}, in \emph{\bibinfo{booktitle}{Adv Neur In}}
  (\bibinfo{year}{2007}).

\bibitem[{\citenamefont{Ing et~al.}(2010)\citenamefont{Ing, Wilson, and
  Geisler}}]{Ing_Wilson_Geisler_2010_JOV}
\bibinfo{author}{\bibfnamefont{A.~D.} \bibnamefont{Ing}},
  \bibinfo{author}{\bibfnamefont{A.~J.} \bibnamefont{Wilson}},
  \bibnamefont{and} \bibinfo{author}{\bibfnamefont{W.~S.}
  \bibnamefont{Geisler}}, \bibinfo{journal}{J Vision}
  \textbf{\bibinfo{volume}{10}}, \bibinfo{pages}{1} (\bibinfo{year}{2010}).

\bibitem[{\citenamefont{Yao et~al.}(2007)\citenamefont{Yao, Shi, Han, Gao, and
  Dan}}]{Yao_Shi_Han_Gao_Dan_2007_NatNeurosci}
\bibinfo{author}{\bibfnamefont{H.}~\bibnamefont{Yao}},
  \bibinfo{author}{\bibfnamefont{L.}~\bibnamefont{Shi}},
  \bibinfo{author}{\bibfnamefont{F.}~\bibnamefont{Han}},
  \bibinfo{author}{\bibfnamefont{H.}~\bibnamefont{Gao}}, \bibnamefont{and}
  \bibinfo{author}{\bibfnamefont{Y.}~\bibnamefont{Dan}}, \bibinfo{journal}{Nat.
  Neurosci.} \textbf{\bibinfo{volume}{10}}, \bibinfo{pages}{772}
  (\bibinfo{year}{2007}).

\bibitem[{\citenamefont{Hua et~al.}(2010)\citenamefont{Hua, Bao, Huang, Wang,
  Xu, Zhou, and Lu}}]{Hua_Bao_Huang_Wang_Xu_Zhou_Lu_2010_CurrBio}
\bibinfo{author}{\bibfnamefont{T.}~\bibnamefont{Hua}},
  \bibinfo{author}{\bibfnamefont{P.}~\bibnamefont{Bao}},
  \bibinfo{author}{\bibfnamefont{C.-B.} \bibnamefont{Huang}},
  \bibinfo{author}{\bibfnamefont{Z.}~\bibnamefont{Wang}},
  \bibinfo{author}{\bibfnamefont{J.}~\bibnamefont{Xu}},
  \bibinfo{author}{\bibfnamefont{Y.}~\bibnamefont{Zhou}}, \bibnamefont{and}
  \bibinfo{author}{\bibfnamefont{Z.-L.} \bibnamefont{Lu}},
  \bibinfo{journal}{Curr Biol} \textbf{\bibinfo{volume}{20}},
  \bibinfo{pages}{887} (\bibinfo{year}{2010}).

\bibitem[{\citenamefont{Li and Gilbert}(2002)}]{Li_Gilbert_2002_JNeurophysiol}
\bibinfo{author}{\bibfnamefont{W.}~\bibnamefont{Li}} \bibnamefont{and}
  \bibinfo{author}{\bibfnamefont{C.~D.} \bibnamefont{Gilbert}},
  \bibinfo{journal}{J. Neurophysiol.} \textbf{\bibinfo{volume}{88}},
  \bibinfo{pages}{2846–2856} (\bibinfo{year}{2002}).

\bibitem[{\citenamefont{Knoblauch and Sommer}(2004)}]{Knoblauch_Sommer_2004}
\bibinfo{author}{\bibfnamefont{A.}~\bibnamefont{Knoblauch}} \bibnamefont{and}
  \bibinfo{author}{\bibfnamefont{F.~T.} \bibnamefont{Sommer}},
  \bibinfo{journal}{Neurocomputing} \textbf{\bibinfo{volume}{58-60}},
  \bibinfo{pages}{185} (\bibinfo{year}{2004}).

\bibitem[{\citenamefont{Hoyer and
  Hyv{\"a}rinen}(2002)}]{Hoyer_Hyvarinen_2002_VisionRes}
\bibinfo{author}{\bibfnamefont{P.~O.} \bibnamefont{Hoyer}} \bibnamefont{and}
  \bibinfo{author}{\bibfnamefont{A.}~\bibnamefont{Hyv{\"a}rinen}},
  \bibinfo{journal}{Vision Res} \textbf{\bibinfo{volume}{42}},
  \bibinfo{pages}{1593} (\bibinfo{year}{2002}).

\bibitem[{\citenamefont{Song et~al.}(2000)\citenamefont{Song, Miller, and
  Abbott}}]{Song_Abbott_Miller_2000}
\bibinfo{author}{\bibfnamefont{S.}~\bibnamefont{Song}},
  \bibinfo{author}{\bibfnamefont{K.~E.} \bibnamefont{Miller}},
  \bibnamefont{and} \bibinfo{author}{\bibfnamefont{L.~F.}
  \bibnamefont{Abbott}}, \bibinfo{journal}{Nat Neurosci}
  \textbf{\bibinfo{volume}{3}}, \bibinfo{pages}{919} (\bibinfo{year}{2000}).

\bibitem[{\citenamefont{Li}(2002)}]{Li_2002}
\bibinfo{author}{\bibfnamefont{Z.}~\bibnamefont{Li}}, \bibinfo{journal}{Trends
  Cogn Sci} \textbf{\bibinfo{volume}{6}}, \bibinfo{pages}{9 }
  (\bibinfo{year}{2002}), ISSN \bibinfo{issn}{1364-6613}.

\bibitem[{\citenamefont{Fukushima}(1980)}]{Fukushima_1980}
\bibinfo{author}{\bibfnamefont{K.}~\bibnamefont{Fukushima}},
  \bibinfo{journal}{Biol Cybern} \textbf{\bibinfo{volume}{36}},
  \bibinfo{pages}{193} (\bibinfo{year}{1980}).

\bibitem[{\citenamefont{Gilbert and Sigman}(2007)}]{Gilbert_Sigman_2007}
\bibinfo{author}{\bibfnamefont{C.~D.} \bibnamefont{Gilbert}} \bibnamefont{and}
  \bibinfo{author}{\bibfnamefont{M.}~\bibnamefont{Sigman}},
  \bibinfo{journal}{Neuron} \textbf{\bibinfo{volume}{54}}, \bibinfo{pages}{667}
  (\bibinfo{year}{2007}).

\bibitem[{\citenamefont{Schneeweis and
  Schnapf}(1995)}]{Schneeweis_Schnapf_1995}
\bibinfo{author}{\bibfnamefont{D.}~\bibnamefont{Schneeweis}} \bibnamefont{and}
  \bibinfo{author}{\bibfnamefont{J.}~\bibnamefont{Schnapf}},
  \bibinfo{journal}{Science} \textbf{\bibinfo{volume}{268}},
  \bibinfo{pages}{1053} (\bibinfo{year}{1995}).

\bibitem[{\citenamefont{Enns and Lollo}(2000)}]{Enns_DiLollo_2000}
\bibinfo{author}{\bibfnamefont{J.~T.} \bibnamefont{Enns}} \bibnamefont{and}
  \bibinfo{author}{\bibfnamefont{V.~D.} \bibnamefont{Lollo}},
  \bibinfo{journal}{Trends Cogn Sci} \textbf{\bibinfo{volume}{4}},
  \bibinfo{pages}{345 } (\bibinfo{year}{2000}), ISSN \bibinfo{issn}{1364-6613}.

\bibitem[{\citenamefont{Fei-Fei et~al.}(2004)\citenamefont{Fei-Fei, Fergus, and
  Perona}}]{FeiFei_2004}
\bibinfo{author}{\bibfnamefont{L.}~\bibnamefont{Fei-Fei}},
  \bibinfo{author}{\bibfnamefont{R.}~\bibnamefont{Fergus}}, \bibnamefont{and}
  \bibinfo{author}{\bibfnamefont{P.}~\bibnamefont{Perona}}, in
  \emph{\bibinfo{booktitle}{CVPR 2004, Workshop on Generative-Model Based
  Vision}} (\bibinfo{year}{2004}).

\bibitem[{\citenamefont{LeCun et~al.}(1998)\citenamefont{LeCun, Bottou, Bengio,
  and Haffner}}]{LeCun_1998}
\bibinfo{author}{\bibfnamefont{Y.}~\bibnamefont{LeCun}},
  \bibinfo{author}{\bibfnamefont{L.}~\bibnamefont{Bottou}},
  \bibinfo{author}{\bibfnamefont{Y.}~\bibnamefont{Bengio}}, \bibnamefont{and}
  \bibinfo{author}{\bibfnamefont{P.}~\bibnamefont{Haffner}}, in
  \emph{\bibinfo{booktitle}{P IEEE}} (\bibinfo{year}{1998}),
  vol.~\bibinfo{volume}{86}, p. \bibinfo{pages}{2278}.

\bibitem[{\citenamefont{Brainard}(1997)}]{Brainard_1997}
\bibinfo{author}{\bibfnamefont{D.~H.} \bibnamefont{Brainard}},
  \bibinfo{journal}{Spatial Vision} \textbf{\bibinfo{volume}{10}},
  \bibinfo{pages}{433} (\bibinfo{year}{1997}).

\end{thebibliography}

\end{document}